\documentclass[a4paper,12pt]{article}
\usepackage{epsfig,float}
\usepackage{cite}
\usepackage[left=2cm,right=2cm]{geometry}
\usepackage{color}
\usepackage{a4wide,graphicx}
\usepackage{morefloats}
\newcommand\T{\rule{0pt}{2.7ex}}
\newcommand\B{\rule[-1.1ex]{0pt}{0pt}}
\newcommand{\eps}{\epsilon}

\begin{document}

\title{Decays of doubly charmed meson molecules}

\author{R. Molina $^1$, H. Nagahiro$^2$ and A. Hosaka$^1$}
\maketitle
\date{}

\begin{center}
$^1$  Research Center for Nuclear Physics (RCNP), Osaka University, Ibaraki, Osaka 567-0047, Japan\\
$^2$ Department of Physics, Nara Women's University, Nara 630-8506, Japan\\
\end{center}

\begin{abstract}  
Several observed states close to the $D\bar{D}^*$ and $D^*_{(s)}\bar{D}^*_{(s)}$ thresholds, as the X(3872) and some XYZ particles can be described in terms of a two-meson molecule. Furthermore, doubly charmed states are also predicted. These new states are near the $D^*D^*$ and $D^*D^*_s$ thresholds, % Therefore, if the previous XYZ are molecules, then, there should be doubly charmed mesons with $J^P=1^+$ around the $D^*D^*$ threshold. 
 %For this reason, it is important to evaluate observables related to them. Because of the spin$=1$, they do not decay into $DD$. In this article we compute decays to $DD^*$ and radiative decays of doubly charmed meson molecules into $DD_{(s)}\gamma$.
and have spin-parity $J^P=1^+$. Their natural decay modes are $D_{(s)}D^*$, $DD_{(s)}\pi$ and $DD_{(s)}\gamma$ and $D^*D_{(s)}\gamma$. We evaluate the widths of these states, named here as $R_{cc}(3970)$ and $S_{cc}(4100)$, and obtain $44$ MeV for the non-strangeness, and $24$ MeV for the doubly charm-strange state. Essentially, the decay modes are $DD_{(s)}\pi$ and $DD_{(s)}\gamma$, being the $D\pi$ and $D\gamma$ emitted by one of the $D^*$ meson which forms the molecule.
\end{abstract}

\section{Introduction}
Since the discovery of the X(3872) by the Belle Collaboration \cite{belle2003}, and observed afterwards by other collaborations, the number of so-called XYZ states have been increasing. The next by mass are X(3915), X(3940), Y(3940), Z(3930) and X(4160). %Although it is not discarded yet some of them could be signs of different decay modes of the same particle, for example, the X(3915) and the Z(3930) \cite{pdg}. 
Most of these states cannot be fitted in terms of $c\bar{c}$ \cite{olsen}. For example, it is difficult to accomodate the X(3872) in the charmonium spectra because the ratio $\Gamma(X\to \pi^+\pi^- J/\psi)/\Gamma(X\to \pi^0\pi^+\pi^-)$ gives large branching fraction to $\pi^+\pi^- J/\psi$, which is unexpected for the available charmonium assignments, $\chi\rq{}_{c1}$ ($2 \,^3P_1$ $c\bar{c}$) or $2^{-+}$, $\eta_{c2}$. % Another reason is that if it was a $2^{-+}$ $c\bar{c}$ state, the decay  $X(3872)\to \gamma J/\psi$ would be hightly suppressed high multipole, which is not the case. 
 Instead, it is often said that the X(3872) could be a molecule made of $D\bar{D}^*$ \cite{Close1,Tornqvist}, which is strongly reinforced by the mass so-close to these thresholds. A clear explanation of this issue has been given in \cite{daniaxial,juandani}, where, based on a molecular picture with an interaction provided by hidden gauge lagrangians \cite{bando,hideko}, the authors found that it is possible to explain the ratio $\Gamma(X\to \pi^+\pi^- J/\psi)/\Gamma(X\to \pi^0\pi^+\pi^-J/\psi)$  as a molecular object of $I=0$ with a mixture of a very small component of $I=1$. Moreover, to explain the missing charged partner and $\chi_{c1}(2P)$, and some reaction rates such as the production from the $p \bar p$ collision, a hybrid picture with a $\bar c c$-core was also proposed \cite{Takizawa}.  While a small fraction of the quark core is needed for the above features, it was shown that the main component of the $X(3872)$ is the $D\bar{D}^*$ molecule. Recently, the LHCb has measured \cite{lhcb2013} the quantum numbers of the X(3872) as $1^{++}$. This result rules out  the X(3872) to be the $\eta_{c2}(1^1 D_2)$ state, favoring the exotic interpretation.  %Taking into account the small mass difference between $D^+$ and $D^0$, together with the large width of the $\rho$-meson compared to the $\omega$, one can explain this ratio implementing the $\rho$-mass distribution in the calculation. Still, the couplings of the X(3872) to $D^+D^{*-}-c.c.$ and $D^0\bar{D}^{*0}-c.c.$ are very similar. Thus, due to the small binding energy in the $D^0\bar{D}^{*0}-c.c.$ channel, the wave function corresponding to this channel extends much further in the space than for $D^+D^{*-}-c.c.$, and the integrated probability of finding the  $D^0\bar{D}^{*0}-c.c.$ component in the wave function is larger \cite{juandani}.
% but in the short range processes that take place close to the origin, both components have similar weight \cite{juandani}.
 %In \cite{aceti}, rates of $J/\psi \omega$ and $J/\psi \rho$, and the ratio $\Gamma(X\to J/\psi \gamma)/\Gamma(J/\psi \pi\pi)$ are evaluated in the molecular picture, finding good agreement with the experiment.  

It has been discussed by several authors whether or not some of the other observed XYZ particles can be described in terms of molecules \cite{xyz,tanja,pavon,guo,LiuLuo,ding,ding2,martinez,segovia,wang}.  Regarding the observed states near to the $D^*\bar{D}^*$ thresholds, using the same formalism as in \cite{daniaxial} for the X(3872), with hidden gauge Lagrangians, some of the XYZ states \cite{olsen} can also be interpreted as $D^*_{(s)}\bar{D}^*_{(s)}$-like molecules \cite{xyz}. In \cite{xyz}, three states with $I=0$ and $J^{PC}=0^{++},1^{+-}$ and $2^{++}$ around $3940$ MeV are found, other with $I[J^{PC}]=0[2^{++}]$ and mass close to $4160$ MeV, and also a $Z_c$ charged state with mass around $3920$ MeV and quantum numbers $I[J^{PC}]=1[2^{++}]$ (see next section for further discussion). Furthermore, in \cite{xiao} the authors show that the dynamics from the extrapolation of the local hidden gauge model to SU(4) fully respects the constraints of heavy quark spin symmetry. In \cite{tanja}, the authors start from the assumption that the Y(3940) and the Y(4140) are hadronic molecules with quantum numbers $J^{PC}=0^{++}$ or $2^{++}$ whose constituents are the charm vectors $D^*\bar{D}^*$ for the Y(3940) and $D_s^{*+}\bar{D}_s^{*-}$ for the Y(4140) and they calculate the decay rates of the observed modes $Y(3940)\to J/\psi \omega$ and $Y(4140)\to J/\psi\phi$ for the case $J^{PC}=0^{++}$. Their results for the coupling constants are the input to evaluate the decay modes that supports the molecular interpretation. In \cite{pavon}, the authors present an EFT (Effective Field Theory) description of heavy mesons molecules based on HQSS (Heavy Quark Spin Symmetry), and predict in total six $D^{(*)}\bar{D}^{(*)}$ molecular states based on the assumption that the X(3915) is a $0^{++}$ heavy spin symmetry partner of the X(3872).

The molecular picture can also be applied to further states, though the vector-vector interaction does not always provide sufficient attraction.  
Only those carrying the quantum number {\it{hidden charm}} $(C=0;S=0)$, $(C=1;S=-1,0,1)$ and $(C=2;S=0,1)$,  form molecules of two $D^{*}$ mesons or $D^*$ and $\rho, K^*$ \cite{Molinarhod, exotic}. Some of these cases are flavour exotics since they can not be reached by $q \bar q$ \cite{exotic}. Of particular interest is the doubly charmed states of $C=2$, which is a challenge also for experiments. Doubly charm states with the same quantum numbers have also been found in \cite{valcarce} from solving the scattering problem of two $D$-mesons with the interaction provided by the chiral constituent quark model. Theoretically, tetraquark structure has been also discussed \cite{cui,marina,Hyodo:2012}. In \cite{cui}, the authors assume that the $X(3872)$ is a tetraquark structure and use its mass as input to determine the mass of the $T_{cc}$, finding $M_{T_{cc}}=3966$ MeV.  Whether they exist or not, and if so what their structure would be like, either molecular or compact tetraquark like, is an important question.  

%The interaction provided by the hidden gauge lagrangian for the vector - vector system, does not predict states in all sectors, only there is enough attraction to bind the system in the sectors $(C=0;S=0)$, $(C=1;S=-1,0,1)$ and $(C=2;S=0,1)$ \cite{exotic}. 
%In \cite{exotic}, some of the difficulties in the quark model do not exist in the molecular picture, like, for instance, the narrow states observed in the first excited state of $D_s$ spectrum, can not be explained by the quark model.   
 %In \cite{exotic}, it is shown that the difficulties that the quark model face, for example, in the first excited state of the $D_s$ spectrum \cite{Godfrey2}, where there are two doublyts of mesons, one with $J^P=0^+,1^+$ and the other with $J^P=1^+,2^+$, as the narrow states with lower mass than the $c\bar{s}$ expectation of the $J^P=0^+,1^+$ doublyt (the observed $D_{s0}(2317)$ and $D_{s1}(2460)$), are fit in the molecular picture, explaining masses and widths of the two doublyts.
On the whole, using the hidden gauge Lagrangians together with unitary scattering amplitudes, it is possible to obtain, the doubly charmed $R_{cc}^+(3970)$ and  $S_{cc}^{+(+)}(4100)$ as dynamically generated states from the vector-vector interaction with the same value of the free parameters (the subtraction constant or cutoff in the two-meson function loop) to those used for the XYZ \cite{xyz} (this is shown in the next section). %(the $\alpha$ of the two-meson loop function in \cite{xyz} is different from the one in \cite{danibreak}, but the results of \cite{xyz} can be obtained by using the same value of $\alpha$ and $\mu$ than in \cite{danibreak, exotic}). 
%Therefore, we can extract the conclusion that if the X(3872), and XYZ in \cite{xyz} are molecules, there should be observed doubly charmed states with $J^P=1^+$ close to the $D^*D^*_{(s)}$ threshold. Thus, it is worth computing more observables on these doubly charmed meson molecules. %One difficulty that we face is the spin of the object, 
 %Since they have $J^P=1^+$, they do not decay into $D\bar{D}$, and therefore further decays should be considered. In this paper, we study decays into $DD^*$ and radiative decays into $DD\gamma$.
 Therefore, if some of these states, like the Y(3940) or X(4160), are good candidates of meson-molecular states,  we strongly expect that there also should be doubly charmed mesons.
 
 To explore further the internal structure of these states, it is useful to study various transition amplitudes, including production and decays.  
In this paper, we study the decays of the predicted exotic states of $D^*D^*_{(s)}$ molecules in detail.  
The production mechanism perhaps requires combinations of hard and soft processes which is beyond our scope in the present paper.  
Because of a bosonic system, a $D^* D^*$ molecule can form a state of spin and parity $J^P = 1^+$ when they are dominated by an s-wave state. They cannot decay into $D\bar{D}$ due to its quantum numbers. In this paper we consider possible decays of $R_{cc}(3970)$ and $S_{cc}(4100)$, including strong and radiative decays. The strong decays occur through $DD^{*}_{(s)}$ or $D_{(s)}D^{*}$ which subsequently go to three body states via $D^* \to \pi D$.  Radiative decays also occur through the above two-body channels with $D^* \to D \gamma$.% and we also evaluate the direct decay processes to $DD^*\gamma$ and $DD\gamma$.  
 There are also direct decays into three-body states, $DD\gamma$, but they are small as compared to the above processes going through two bodies, as we show in this manuscript.

%It is important to study more physical observables for the new exotic particles which are useful to get information for the internal structure of  possible observed states. Because of a bosonic system, a $D^* D^*_{(s)}$ molecule can form a state of spin and parity $J^P = 1^+$ when they are dominated by an S-wave state.  In this paper we consider the possible decays of the $R_{cc}^+(3970)$, $S_{cc}^{+}(4100)$ and  $S_{cc}^{++}(4100)$, which are essentially $D^0D^{*+}_{(s)}$, $D^{*0}D^{+}_{(s)}$ and $D^{*+}D^{+}_s$, which subsequently goes to three body states via $D^*\to D\pi$. There are also direct decays into three-body states, but they are small as compared to the above processes going through two bodies.
The structure of the paper is as follows. First, we explain briefly the model for the dynamically generated resonances, XYZ and doubly charm states, in section 2. In section 3, we evaluate their decay widths to $D_{(s)}D^*_{(s)}$, which are mediated by vector mesons or pseudoscalar ones. Direct three-body radiative decays are studied in section 4 and 5. Finally, in sections 6 we show the results reaching the conclusions and final remarks in section 7.
\section{Dynamically generated XYZ, doubly charm mesons and the breaking of the SU(4) symmetry}

\begin{figure}
\begin{center}
\includegraphics[width=12cm]{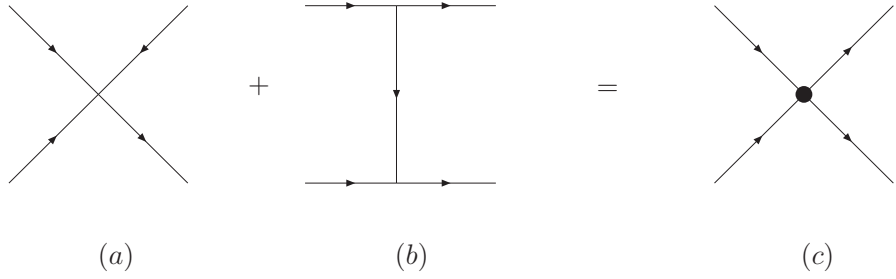}
\end{center}
\caption{Feynman diagrams for the vector - vector  interaction, contact term (a) and vector exchange (b) are included in the kernel $V$\cite{exotic}.}
\label{fig:fig1} 
\end{figure}

\begin{figure}
\begin{center}
\includegraphics[width=12cm]{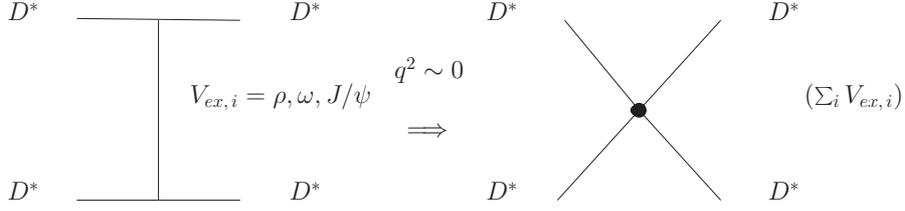}
\end{center}
\caption{Point-like vector-vector interaction in the case of $D^*D^*\to D^*D^*$.}
\label{fig:q2} 
\end{figure}

In our approach, local hidden gauge lagrangians are used to study the vector-vector interaction \cite{raquel1,geng, Molinarhod,xyz,exotic}. The hidden gauge Lagrangian, which involves the interaction of 
vector mesons amongst themselves, coming from the formalism of the hidden gauge symmetry (HGS) for vector mesons \cite{bando,hideko}
\begin{equation}
{\cal L}_{III}=-\frac{1}{4}\langle V_{\mu \nu}V^{\mu\nu}\rangle \ ,
\label{lVV}
\end{equation}
where the symbol $\langle \rangle$ stands for the trace in the $SU(4)$ space 
and $V_{\mu\nu}$ is given by 
\begin{equation}
V_{\mu\nu}=\partial_{\mu} V_\nu -\partial_\nu V_\mu -ig[V_\mu,V_\nu]\ ,
\label{Vmunu}
\end{equation}
with $g$ given by
\begin{equation}
g=\frac{M_V}{2f}\ ,
\label{g}
\end{equation}
and $f=93$ MeV the pion decay constant. Using the value of $g$ in Eq.~(\ref{g}) is 
one of the ways to account for the KSFR relation \cite{KSFR} which
is tied to  
vector meson dominance \cite{sakurai}.
% The magnitude $V_\mu$ is the $SU(4)$ 
%matrix of the vectors of the 15th representation together with the singlet to
%implement the mixing of the $\phi$, $\omega$ and $J/\psi$, which is
The vector field $V_\mu$ is represented by the $SU(4)$ matrix which is
parametrized by 16 vector mesons including the 15-plet and singlet of $SU(4)$,
\begin{equation}
V_\mu=\left(
\begin{array}{cccc}
\frac{\rho^0}{\sqrt{2}}+\frac{\omega}{\sqrt{2}}&\rho^+& K^{*+}&\bar{D}^{*0}\\
\rho^-& -\frac{\rho^0}{\sqrt{2}}+\frac{\omega}{\sqrt{2}}&K^{*0}&D^{*-}\\
K^{*-}& \bar{K}^{*0}&\phi&D^{*-}_s\\
D^{*0}&D^{*+}&D^{*+}_s&J/\psi\\
\end{array}
\right)_\mu \ ,
\label{Vmu}
\end{equation}
where the ideal mixing has been taken for $\omega$, $\phi$ and $J/\psi$.
The interaction of ${\cal L}_{III}$ gives rise to a contact term 
\begin{equation}
{\cal L}^{(c)}_{III}=\frac{g^2}{2}\langle V_\mu V_\nu V^\mu V^\nu-V_\nu V_\mu
V^\mu V^\nu\rangle\ ,
\label{lcont}
\end{equation}
depicted in Fig.~\ref{fig:fig1} (a), and also to a three 
vector vertex \cite{raquel1,geng}
\begin{equation}
\mathcal{L}_{3V}=ig\langle (V^\mu \partial_\nu V_\mu -\partial_\nu V_\mu V^\mu)
V^\nu)\rangle\ .
\label{l3V}
\end{equation}
This latter Lagrangian gives rise to a
$VV\to VV$ amplitude by means of the exchange of one of the vectors, as 
shown in Figs.~\ref{fig:fig1} (b).
The vector-exchange diagram in Fig. \ref{fig:fig1} (b) dominates the interaction (in the sectors $charm=2;strangeness=0,1$, only the potential from this diagram survives) and upon the aproximation $q^2\sim 0$ leads to a contact interaction, see Fig. \ref{fig:q2}. In some channels it is attractive and leads to the generation of bound states in the coupled channel calculation \cite{xyz}. The potential from Fig. \ref{fig:fig1} (b) after spin projection can be read as:
\begin{eqnarray}
V_{ij}=&-&\frac{g^2}{m^2_{V_{ex}}}A_{ij}(s-u) \quad \mathrm{for}\quad J=0,1,2\ , \nonumber\\
\end{eqnarray}
in the $t$-channel and
\begin{eqnarray}
V_{ij}=&-&\frac{g^2}{m^2_{V_{ex}}}B_{ij}(s-t) \quad \mathrm{for} \quad J=0,2\ , \nonumber\\
V_{ij}=&&\frac{g^2}{m^2_{V_{ex}}}B_{ij}(s-t) \quad \mathrm{for} \quad J=1\ ,
\label{eq:exchp}
\end{eqnarray}
in the $u$-channel. Here, $m_{V_{ex}}$ stands for the mass of the exchanged vector meson, $A_{ij},B_{ij}$ are the coefficients for the particular transitions $i \to$ $j$, and $g=m_V/(2f)$.

\begin{figure}
\begin{center}
\includegraphics[width=8cm]{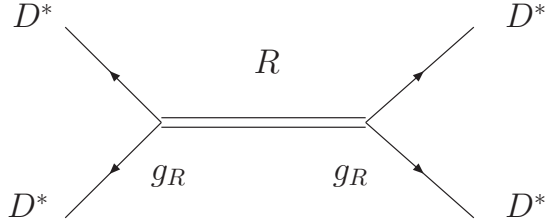}
\end{center}
\caption{Couplings of the resonance to the two vector meson component.}
\label{fig:2} 
\end{figure}

The $SU(4)$ structure of the Lagrangian allows us to take into account all possible particle
channels by a single interaction term. However, in reality, the SU(4) flavour symmetry is broken in several ways:

\begin{itemize}
\item[1)] All the physical masses are taken from the PDG \cite{pdg}. The strength of the $s$-wave projected potential of Eq. (\ref{eq:exchp}) increases with the initial energy $s$, which leads to a stronger potential when $D^*$ particles are involved than for lighter vector mesons. 
\item[2)] The exchange of heavy particles like $J/\psi$ is supressed compared to $\rho, \omega$ or $\phi$ due to the presence of $m_{V_{ex}}^2$ in the denominator. 
\item[3)] Different coupling constants \lq\lq{}$g$\rq\rq{} in Eqs. (\ref{lcont}), (\ref{l3V}) and (\ref{eq:exchp}) are taken, \lq\lq{}$g_h^2$\rq\rq{} for $V_h V_h \to V_h V_h$, \lq\lq{}$g_h g_l$\rq\rq{} for $V_h V_h \to V_l V_l$, and \lq\lq{}$g_l^2$\rq\rq{} for $V_l V_l \to V_l V_l$, where $h$ means heavy particle, $h=D^*,D^*_s$ and $l$ light particle. Thus different coupling constants are used $g_{D^*}=m_{D^*}/(2f_D)$, $g_{D^*_s}$ or $g=m_\rho/(2f_\pi)$, with $f_D=206/\sqrt{2}$ MeV and $f_\pi=93$ MeV in \cite{xyz}.
\end{itemize}
The sum of the amplitudes in Fig. \ref{fig:fig1}, projected in $s$-wave, isospin and spin, are the kernel of the Bethe Salpeter equation used to unitarize amplitudes resumming over loops \cite{oller}. 
\begin{equation}
T= (\hat{1}-VG)^{-1} V\ .
\label{Bethe}
\end{equation}
 The potencial $V$ here is a $10\times 10$ matrix in $I=0$ with the amplitudes obtained from the channels 
$D^*\bar{D}^*$, $D^*_s\bar{D}^*_s$, $K^*\bar{K}^*$, $\rho\rho$, $\omega\omega$, $\phi\phi$, $J/\psi J/\psi$, $\omega J/\psi$, $\phi J/\psi$, $\omega \phi$, in its elements for each spin $J=0,1,2$ independently. Also, in $I=1$, $V$ is a $6\times6$ matrix whose elements come from the transition potentials between the channels $D^*\bar{D}^*$, $K^*\bar{K}^*$, $\rho\rho$, $\rho\omega$, $\rho J\psi$, $\rho\phi$, and $G$ is a diagonal matrix where its elements are the two meson loop 
function $G_i(P)$ for each channel $i$
\begin{equation}
G_i(P)=i\int \frac{d^4 q}{(2\pi)^4}\frac{1}{q^2-m_1^2+i\eps}\frac{1}{(P-q)^2-m_2^2+i\eps}\ ,
\label{loop}
\end{equation}
where $P$ is the initial four-momentum, with $P^0=\sqrt{s}$. $G_i(P)$ is a function of $\alpha(\mu)$ in the dimensional regularization scheme, or $q_{\mathrm{max}}$ if the cutoff method is used instead. The result of the calculation gives the poles positions in the complex plain and the coupling constant of the state, $g_R$, to the two meson component in each channel, evaluated as the square root of the residue of the pole in this channel, see Fig. \ref{fig:2}. This coupling constant $g_R$ and the pole positions, $\sqrt{s_0}$, are the only magnitudes needed to evaluate observables as decay widths.

To investigate the effect of the SU(4) breaking, we have performed the calculation of \cite{xyz} by using two parameter sets: 1) with SU(4) symmetric coupling, $g_l=m_\rho/(2f_\pi)$ in Eqs. (\ref{lcont}), (\ref{l3V}) and $\alpha_h=-1.4$ ($\mu=1500$ MeV) in all channels, and 2) with SU(4) breaking couplings, we use $g^2_h$ or $g_lg_h$, being $g_h=m_{D^*}/(2f_D)$ in the channels involving heavy particles as explained before in this section, and $\alpha_h=-1.27$. The results are shown in the Tables \ref{xyza} and \ref{xyzb} in the Appendix (A.1), where pole positions, $\sqrt{s_0}$, and couplings to the two-meson channel of the resonance, $g_R$ are summarized. From these tables, we can see that: 1) the use of different $g_l$ or $g_h$ in the heavy channels is mostly compensated by a small change of $\alpha_h$. 2) the value of the coupling of the resonance, $g_R$, to the most important channel, $D^*\bar{D}^*$ or $D^*_{s}\bar{D}^*_{s}$, barely changes using one or the other prescription.
The conclusion is that the different value of the coupling $g$ in SU(3), for heavy particles, can be absorved into the change in the subtraction constant $\alpha_h$.
%Using these two sets of parameters we evaluate the pole positions and couplings of the resonance for the doubly charm states. The results are summarized in Table \ref{xyzdo}. The conclusion is that the different value of the coupling $g$ in SU(3), for heavy particles, can be absorved into the change in the subtraction constant $\alpha_h$.%This coupling $g_R$ is the quantity needed to evaluate the decay widths of the bound state as we show in section 3?.

 In this article we refer to those states obtained in \cite{exotic} in the sectors $charm=2; strangeness=0,1$, where only one channel is possible, $D^*D^*$ or $D^*D^*_s$, as $R_{cc}(3970)$ and $S_{cc}(4100)$. These bound states have quantum numbers $I[J^P]=0[1^+]$ and $I[J^P]=1/2[1^+]$, respectively.

 The amplitude in Fig. \ref{fig:2} shows diagrammatically the coupling of the $R_{cc}$ ($S_{cc}$) resonance to $D^*D^*$, $D^*D^*_s$. For the couplings of the resonance to the two vectors, keeping the $spin=1$ structure, the resulting $T$-matrix amplitude $T(D^*D^*\to D^*D^*)$ can be approximated in the vecinity of the pole as 
\begin{equation}
T\simeq\frac{[g_{R}\frac{1}{2}(\eps^i_1\eps^j_2-\eps^j_1\eps^i_2)][g_{R}\frac{1}{2}(\eps^i_1\eps^j_2-\eps^j_2\eps^i_2)]}{s-s_p}
\end{equation}

Using the two sets of parameters, 1) $g_l$ with $\alpha_h=-1.4$, or 2) $g_h$ with $\alpha_h=-1.27$, to obtain the $I=0$ states around $3940$ MeV as described above, we evaluate the pole positions and couplings of the resonance for the doubly charm states. The results are summarized in Table \ref{xyzdo}. In this table we show the value of the coupling to the most important channel comparing both assumptions. As we observe under these two different situations,  the changes in the masses and coupling constants are small, in particular the resulting coupling constants are similar to those obtained  by the Weinberg\rq{}s formula,
\begin{equation}
\frac{g^2_W}{4\pi}=4(m_1+m_2)^2\sqrt{\frac{2B}{\mu}}[1+O(\frac{\sqrt{2\mu B}}{\beta})]
\end{equation}
where $m_{1,2}$ are the constituent masses, $B$ is the binding energy of the molecule with mass $M$, defined as $M=m_1+m_2-B$, $\mu$ is the reduced mass and $1/\beta$ the range of the forces \cite{wein,guomol}.
 We obtain in both cases doubly charm states. Their couplings to $D^*D^*_{(s)}$ are used to evaluate their decay widths in the next section. 
 
 From Table \ref{xyzdo}, we see, for only one channel as the case of doubly charm mesons ($D^*D^*$ or $D^*D^*_s$), the decrease in the mass reverts into a larger couplings, which is consistent with the formula of Weinberg. 
 
 This effect of bigger coupling for lower masses will be taken into account in this paper when we perform the calculation of the errors of the decay widths.%, by looking for a simple linear relation between the cutoff used in the two meson function loop and the binding energy and coupling of the resonance to $D^*D^*$, then varying these parameteres. 

\begin{table}[htb]
\begin{center}
\begin{tabular}{lrrrrrrrr}
\hline
$C;S$&I,J&\multicolumn{2}{c}{$M_R$}&channel&\multicolumn{2}{c}{$|g_R|$}&\multicolumn{2}{c}{$g_{eff}^{W}$}\T\B\\\hline
&&($g_l$)&($g_h$)&&($g_l$)&($g_h$)&($g_l$)&($g_h$)\T\B\\
$0;0$&$0,0$&$3936$&$3950$&$D^*\bar{D}^*$&$18700$&$18000$&$18050$&$17200$\T\B\\
&$0,1$&$3940$&$3955$&$D^*\bar{D}^*$&$18260$&$17200$&$17800$&$16900$\T\B\\
&$0,2$&$3921$&$3922$&$D^*\bar{D}^*$&$20600$&$21000$&$18800$&$18800$\T\B\\
&$0,2$&$4174$&$4160$&$D^*_s\bar{D}^*_s$&$20400$&$19500$&$16700$&$17700$\T\B\\
&$1,2$&$3970 $&$3924$&$D^*\bar{D}^*$&$20500$&$20560$&$15800$&$18700$\T\B\\
$2;0$&$0,1$&$3968$&$3942$&$D^*D^*$&$16800$&$19500$&$15900$&$17800$\T\B\\
$2;1$&$1/2,1$&$4100$&$4070$&$D^*_sD^*$&$13400$&$17700$&$13100$&$16400$\T\B\\
\hline\hline\end{tabular}
\end{center}
\caption{Mass of the dynamically generated XYZ, doubly charm resonances, and the coupling to the main channel using $g_l$ and $\alpha_h=-1.4$ or $g_h$, with $\alpha_h=-1.27$ in the channels with heavy vector mesons. Units are MeV.}
\label{xyzdo}
\end{table}

\section{Decays of doubly charmed states to $D_{(s)}D^*_{(s)}$}

\begin{figure}
\begin{center}
\includegraphics[width=12cm]{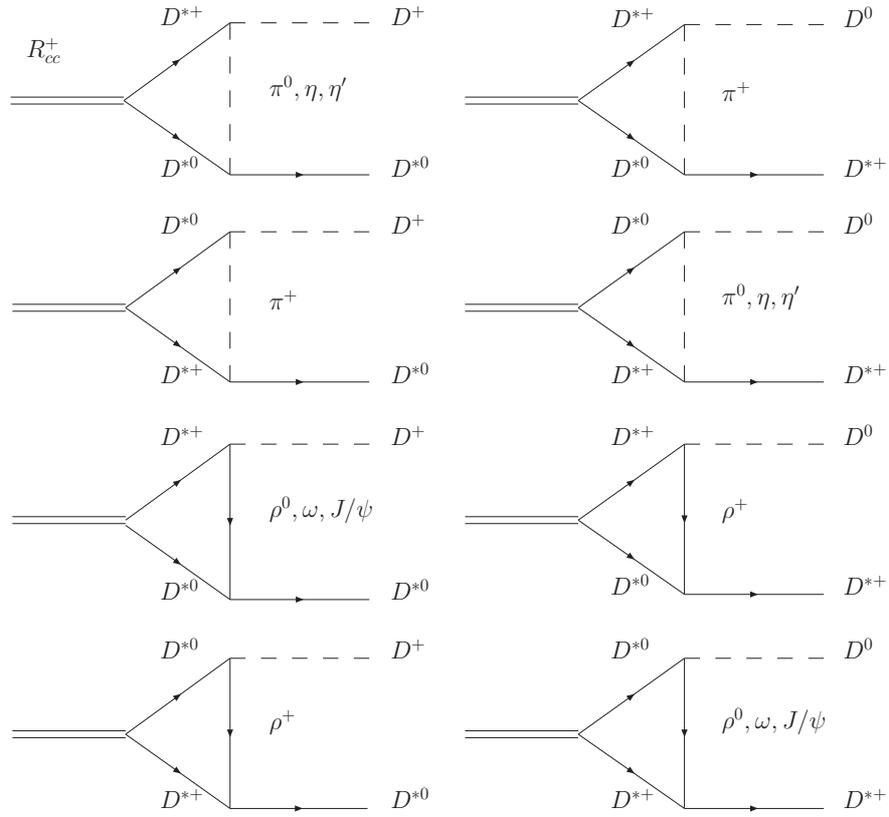}
\end{center}
\caption{Feynman diagrams evaluated in the decay $R_{cc}\to DD^*$. }
\label{fig:3} 
\end{figure}

%The Hidden Gauge Lagrangian \cite{bando,hideko} provides the interactions for vector mesons, among themselves, with pseudoscalars
%and with a photon.% In the coupling of pseudoscalars with photons, the vector meson dominance is implicit since the photon has no other chances that 
%coupling to the vector meson. 
The $R_{cc}\to DD^*_{(s)}$ transition can be reached through anomalous couplings $VVP$ (where the symbol $R_{cc}$ stands for the doubly charmed resonance). The set of Feynman diagrams considered are depicted in Fig. \ref{fig:3}.
 Within loops, particles cannot be distinguished and we must consider different sign of the isospin factor in the isospin combination:
\begin{equation}
\vert D^*D^*,I=0,I_3=0\rangle=\frac{1}{\sqrt{2}}(-D^{*+}D^{*0}+D^{*0}D^{*+})
\end{equation}
with $D^{*+}=\vert \frac{1}{2},\frac{1}{2}\rangle$ and $D^{*0}=-\vert \frac{1}{2},-\frac{1}{2}\rangle$. While for the strange doubly charmed state, we have $\vert D^{*}D^{*}_s,I=\frac{1}{2},I_3=-\frac{1}{2}\rangle=-D^{*0}D^{*+}_s$ and $\vert D^{*}D^{*}_s,I=\frac{1}{2},I_3=\frac{1}{2}\rangle=D^{*+}D^{*+}_s$. The couplings of the resonances to the $D^*D^*$ component are given in Table \ref{xyzdo} in the isospin basis, using two different prescriptions: 1) $\alpha_h=-1.4$ and $g_l=m_\rho/(2f_\pi)$, 2) $\alpha_h=-1.27$ and $g_h=m_{D^*}/(2f_D)$. We will use the first prescription and perform variations of the parameters involved afterwards (see section 6). %vary the couplings, $g$ and $g_R$, binding energy and free parameter of the two-meson function loop consistently in the evaluation of the errors (see section 6). % (that contains implicitly the $\frac{1}{\sqrt{2}}$ factor in the unitary normalization for identical particles) but multiplying by the isospin factor $I=-\frac{1}{\sqrt{2}},\frac{1}{\sqrt{2}}$ for $D^{*+}(q)D^{*0}(P-q)$ or $D^{*0}(q)D^{*+}(P-q)$ respectively. 

The Lagrangians needed to evaluate the decay width to $DD^*_{(s)}$ are \cite{hideko},
\begin{eqnarray}
%\mathcal{L}_{V\gamma}&=&-M_V^2\frac{e}{g}A_\mu\langle V^\mu Q\rangle\label{lag2}\\
&&\mathcal{L}_{PPV}=-ig\langle V^\mu [P,\partial_\mu P]\rangle\label{lag3}\nonumber\\
&&\mathcal{L}_{3V}=ig\langle (V^\mu \partial_\nu V_\mu -\partial_\nu V_\mu V^\mu)
V^\nu)\rangle\label{lag4}\nonumber\\
&&\mathcal{L}_{VVP}=\frac{G'}{\sqrt{2}}\epsilon^{\mu\nu\alpha\beta}\langle\partial_\mu V_\nu \partial_\alpha V_\alpha P\rangle\label{lagan}\ ,
\end{eqnarray}
with $e$ the unit electronic charge ($e^2/4\pi=\alpha$), $G' = 3g'^2/(4\pi^2f)$, $g' = -G_V M_\rho/(\sqrt{2}f^2)$, $G_V=f/\sqrt{2}$ and $g=M_V/2f$. The constant $f$ is the pion decay constant $f=93\ MeV$, $Q=diag(2,-1,-1,1)/3$ and $M_V$ is the mass of the vector meson.%or w, fhich we take $M_{\rho}=770$ MeV. 

The $P$ matrix contain the 15-plet of the pseudoscalars and the 15-plet of vectors respectively in the physical basis considering $\eta$, $\eta'$ mixing \cite{gamphi3770}:%(\textbf{see why the $\eta$ is different of the $\eta_8$ with $\sqrt{6}$})
\begin{equation}
P=\left(
\begin{array}{cccc}
\frac{\eta}{\sqrt{3}}+\frac{\eta'}{\sqrt{6}}+\frac{\pi^0}{\sqrt{2}} & \pi^+ & K^+&\bar{D}^0\\
\pi^- &\frac{\eta}{\sqrt{3}}+\frac{\eta'}{\sqrt{6}}-\frac{\pi^0}{\sqrt{2}} & K^{0}&D^-\\
K^{-} & \bar{K}^{0} &-\frac{\eta}{\sqrt{3}}+\sqrt{\frac{2}{3}}\eta'&D^-_s\\
D^0&D^+&D^+_s&\eta_c
\end{array}
\right)\ ,
\end{equation}
%and $V_\mu$ represents the vector nonet with ideal mixing:
 %\begin{equation}
%\renewcommand{\tabcolsep}{1cm}
%\renewcommand{\arraystretch}{2}
%V_\mu=\left(
%\begin{array}{cccc}
%\frac{\omega+\rho^0}{\sqrt{2}} & \rho^+ & K^{*+}&\bar{D}^{*0}\\
%\rho^- &\frac{\omega-\rho^0}{\sqrt{2}} & K^{*0}&D^{*-}\\
%K^{*-} & \bar{K}^{*0} &\phi&D^{*-}_s\\
%D^{*0}&D^{*+}&D^{*+}_s&J/\psi
%\end{array}
%\right)_\mu\ .
%\end{equation}
\begin{figure}
\begin{center}
\includegraphics[width=13cm]{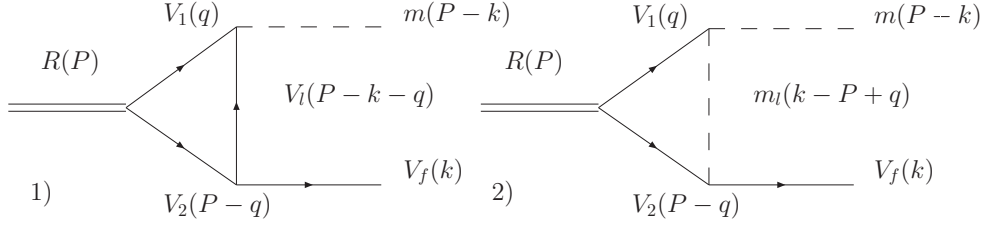}
\end{center}
\caption{PPV and 3V Feynman diagrams for the evaluation of the $R_{cc}\to DD^*_{(s)}$ decay width.}
\label{fig:4} 
\end{figure}

The Feynman diagrams involved in the evaluation of the decay width of $R_{cc}\to DD^*$ are depicted in Fig. \ref{fig:3}. 
They are similar to those of \cite{k2,aceti} since the transition $VV\to VP$ is also there.
Essentially, we have two different structures which are depicted in Fig. \ref{fig:4}. Diagram 1) contains a 3V- vertex and 2) has a PPV- vertex instead. Below, we evaluate both diagrams in Fig. \ref{fig:4}.
\subsection{Decay mediated by a vector meson, Fig. \ref{fig:4}. 1)}
To evaluate diagram 1), first we show the structures of the three different vertices that appear in Fig. \ref{fig:4}, 1) (derived from Eq. (\ref{lagan})):
\begin{eqnarray}
&&t_{RVV}=\frac{I\,g_{R}}{2}(\eps^i_1\eps^j_2-\eps^j _1\eps^i_2)\label{rvv}\\
&&t_{V_3}=V_3 g \lbrace (2k+q-P)^\mu \eps_{(l)\,\nu}\eps_{2\,\mu}\eps^\nu_{(f)}-(k+P-q)^\mu\eps_{2\,\nu}\eps_{(l) \,\mu}\eps^\nu_{(f)}+(2(P-q)-k)_\mu\eps_{(l)\,\nu}\eps^\mu_{(f)}\eps^\nu_2\rbrace\nonumber\\\label{eq:3vertex}\\
\label{eq:ts}
&&t_{VVP}=A G' \eps^{\alpha\beta\gamma\delta} (P-q)_\alpha\eps_{2\,\beta} k_\gamma\eps_{(f)\,\delta}\ .\label{eq:ts}
\end{eqnarray}
Here $P$ is the total (initial) momentum of $R_{cc}$, $k$ is the final momentum of the vector and $q$, the internal loop momentum, as assigned in Fig. \ref{fig:4}. $\eps$ symbols with the indices $1,2$ and $(l),(f)$ are the polarization vectors of the resonance $R=R(1,2)$, of the exchanged vector and the final vector meson respectively. After summing over polarizations,
%\begin{eqnarray}
%t_{V_3}=V_3 g \lbrace (2k+q-P)^\mu \eps_{(l)\,\nu}\eps_{2\,\mu}\eps^\nu_{(f)}-(k+P-q)^\mu\eps_{2\,\nu}\eps_{(l) \,\mu}\eps^\nu_{(f)}+(2(P-q)-k)_\mu\eps_{(l)\,\nu}\eps^\mu_{(f)}\eps^\nu_2\rbrace\label{eq:3vertex}
%\end{eqnarray}
\begin{eqnarray}
&&\sum \eps_{1\,i}\eps_{1\,k}=\delta_{ik}\ ,\nonumber\\
&&\sum \eps_{2\,j}\eps_{2\,l}=\delta_{jl}\ ,\label{eq:sumpol}
\end{eqnarray}
for the two-heavy and near-threshold $D^*_{(s)}$, and,
\begin{eqnarray}
&&\sum \eps_{(l)\,\delta}\eps_{(l)\,\nu}=-g_{\delta\nu}+\frac{(P-k-q)_\delta (P-k-q)_\nu}{M^2_l}\ ,\label{eq:sumpol1}
\end{eqnarray}
for the vector meson exchanged between $V_1$ and $V_2$, hence, the amplitude can be written as,
\begin{eqnarray}
&&-it^{ij}=-\frac{1}{2}AIV_3 g_{R}G\rq{}g\int \frac{d^4q}{(2\pi)^4}\frac{q_\alpha(P-k)_\gamma}{(q^2-m^2_1)((P-q)^2-m^2_2)((k+q-P)^2-M^2_l)}\nonumber\\&&\times\lbrace \eps_{(f)\,\delta}((2k+q-P)^j\eps^{\alpha i\gamma\delta}-(2k+q-P)^i\eps^{\alpha j\gamma\delta}\nonumber\\&& -(k+P)_\delta(\eps^{\alpha i\gamma\delta}\eps^j_{(f)}-\eps^{\alpha j\gamma \delta}\eps^i_{(f)})\nonumber\\&&+(2(P-q)-k)_\mu\eps^\mu_{(f)}(\eps^{\alpha i\gamma j}-\eps^{\alpha j\gamma i})\rbrace\ .\label{eq:5}
\end{eqnarray}
The second term is proportional to $(k_\gamma P_\delta -P_\gamma k_\delta)(\eps^{\alpha i\gamma \delta}\eps^j_{(f)}-\eps^{\alpha j\gamma \delta}\eps^i_{(f)})=-2k_\gamma P_\delta (\eps^{\alpha i \gamma \delta }\eps^j_f-\eps^{\alpha j\gamma \delta }\eps^i_{(f)})$. We apply a Feynman parametrization for this integral,
\begin{eqnarray}
\frac{1}{abc}=2\int^1_0 dx\int^x_0 dy\frac{1}{(a+(b-a) x+(c-b) y)^3}\ ,\label{eq:fein1}
\end{eqnarray}
with 
\begin{eqnarray}
a=q^2-m^2_1; \hspace{0.5cm}b=(P-q)^2-m^2_2;\hspace{0.5cm}c=(P-q-k)^2-M^2_l\ .\label{eq:fein2}
\end{eqnarray}
Using the change of variable $q\rq{}=q-Px+ky$, it can be seen that the second term in Eq. (\ref{eq:5}) is proportional to
\begin{eqnarray}
-2\int^1_0 dx\int^x_0 dy \int  \frac{d^4 q\rq{}}{(2\pi)^4}\frac{(q\rq{}+Px-ky)_\alpha}{(q\rq{}^2+s(M_l))^3} 2k_\gamma P_\delta (\eps^{\alpha i \gamma \delta }\eps^j_f-\eps^{\alpha j\gamma \delta }\eps^i_f)\ ,
\end{eqnarray}
with 
\begin{eqnarray}
s(M_l)=(P^2-m^2_2+m^2_1)x+(-2Pk-M^2_l+m^2_2+k^2) y -(Px-yk)^2-m^2_1\ .\label{eq:fein3}
\end{eqnarray}
The term which is odd in $q\rq{}$ vanishes and those not odd in $q\rq{}$ are proportional either to $k_\gamma P_\delta P_\alpha \eps^{\alpha i \gamma \delta} $ or $k_\gamma P_\delta k_\alpha \eps^{\alpha i \gamma \delta}$, both being equal to zero. Hence, two different kinds of integral, which are parametrized using Lorentz covariance, remain in Eq. (\ref{eq:5}),
\begin{eqnarray}
&&\int \frac{d^4 q}{(2\pi)^4}\frac{q_\alpha (2 k+q-P)_j}{(q^2-m^2_1) ((P-q)^2-m^2_2) ((k+q-P)^2-M^2_l)}\nonumber\\
&&=i(aP_\alpha P_j+bP_\alpha k_j + ck_\alpha P_j+d k_\alpha k_j +e g_{\alpha j})\nonumber\\
&& \int\frac{d^4 q}{(2\pi)^4}\frac{q_\alpha (2(P-q)-k)_\mu}{(q^2-m^2_1) ((P-q)^2-m^2_2) ((k+q-P)^2-M^2_l)}\nonumber\\
&&=i(A P_\alpha P_\mu +BP_\alpha k_\mu+C k_\alpha P_\mu +Dk_\alpha k_\mu +E  g_{\alpha\mu})\label{eq:int3}
\end{eqnarray}
In Eq. (\ref{eq:int3}), for $P_j=0$ as in the center of mass system, only the terms proportional to $b,d$ and $e$ survives. Whereas when contracting with $\eps^\mu_{(f)}$ in the Lorentz gauge, terms with $B$ and $D$ are zero. 

Let us start with the second integral.  For convenience, we separate it into two terms, one that goes like $q_\alpha q_{\mu}$, and the other, proportional to $q_\alpha$, with coefficients $-2$ and $(2 P-k)^{\mu}$, respectively. We define, 
\begin{eqnarray}
&&\int \frac{d^4 q}{(2\pi)^4}\frac{q_\alpha q_{\mu}}{(q^2-m^2_1) ((P-q)^2-m^2_2) ((k+q-P)^2-M^2_l)}\nonumber\\
&&=i(A_1 P_\alpha P_\mu +B_1P_\alpha k_\mu+C_1 k_\alpha P_\mu +D_1k_\alpha k_\mu +E_1 g_{\alpha\mu})\ .\label{eq:int4}
\end{eqnarray}
For simplicity, we work in the center of mass system with the $z$ axes defined in the direction of $\vec{k}$, the momenta of the final vector meson. Thus, we have, for the r. h. s. of Eq. (\ref{eq:int4}),
\begin{itemize}
\item[1)] For $\alpha=0;\mu=0$, $i(A_1 P_0^2 +B_1P_0 k_0+C_1 k_0 P_0 +D_1k_0^2 +E_1  )$,
%\begin{eqnarray}
%&&\int\frac{d^4 q}{(2\pi)^4}\frac{q_0^2}{(q^2-m^2_1) ((P-q)^2-m^2_2) ((k+q-P)^2-M^2_l)}\nonumber\\
%&&\label{eq2:1}
%\end{eqnarray}
\item[2)] For $\alpha=0;\mu=3$, $i(B_1P_0 k_3 +D_1k_0 k_3)$,
%\begin{eqnarray}
%&&\int\frac{d^4 q}{(2\pi)^4}\frac{q_0 q_3}{(q^2-m^2_1) ((P-q)^2-m^2_2) ((k+q-P)^2-M^2_l)}\nonumber\\
%&&\label{eq2:2}
%\end{eqnarray}
\item[3)] For $\alpha=3;\mu=3$, $i(D_1k_3^2 -E_1  )$.
%\begin{eqnarray}
%&&\int\frac{d^4 q}{(2\pi)^4}\frac{q_3^2}{(q^2-m^2_1) ((P-q)^2-m^2_2) ((k+q-P)^2-M^2_l)}\nonumber\\
%&&\ .\label{eq2:3}
%\end{eqnarray}
\end{itemize} 
%The coefficients $B_1$ and $C_1$ are identical since they come from the same integral.
Let us start with the second case of $\alpha=0;\mu=3$, where the integral of the l. h. s. of Eq. (\ref{eq:int4}) is convergent. By taking the change of variable $q\rq{}=q-Px+ky$ and using the same Feynman parametrization of Eqs. (\ref{eq:fein1}), (\ref{eq:fein2}) and (\ref{eq:fein3}), the integral is odd in $q\rq{}_0 $ and/or $q\rq{}_3$, and so we obtain,
\begin{eqnarray}
&&B_1=-\frac{1}{16\pi^2}\int^1_0 dx\int^x_0 dy\frac{xy}{s(M_l)+i\eps};\hspace{0.5cm}C_1=B_1;\hspace{0.5cm}D_1=\frac{1}{16\pi^2}\int^1_0 dx\int^x_0 dy\frac{y^2}{s(M_l)+i\eps}\ ,\nonumber\\
\label{eq:cstes}
\end{eqnarray}
where we have used the relation: $\int d^4 q\rq{}/(q\rq{}^2+s)^3=i \pi^2/(2 s)$. 

Now inserting the coefficients of Eq. (\ref{eq:cstes}) in the resulting equation from Eq. (\ref{eq:int4}) for each case, $\alpha=0,\mu=0$ and $\alpha=3,\mu=3$, we obtain the coefficients $A_1$ and $E_1$, 
\begin{eqnarray}
&&E_1= \frac{i}{(2 \pi)^3}f(P,k,M_l)+\frac{k^2_3}{16 \pi^2} \int^1_0 dx \int^x_0 dy \frac{y^2}{s(M_l)+i \eps}\ ,\nonumber\\\label{eq:e}\\
&&A_1=\frac{1}{P_0^2}\lbrace -\frac{i}{(2 \pi)^3} f_1(P,k,M_l)-\frac{1}{16 \pi^2} \int_0^1dx\int_0^x dy\frac{k^2_3 y^2+k^2_0 y^2-2 P_0 k_0 x y}{s(M_l)+i \eps}\ ,\rbrace\label{eq:a1}
\end{eqnarray}
where
\begin{eqnarray}
&&f(P,k,M_l)=\int  \frac{dq^0\,d\mathrm{cos}\,\theta dqq^4 \mathrm{cos}^2\theta}{(q^2-m^2_1+i\eps) ((P-q)^2-m^2_2+i\eps) ((k+q-P)^2-M^2_l+i\eps)}\label{eq:f}\\
&&f_1(P,k,M_l)=\int \frac{dq^0\,d\mathrm{cos}\,\theta dqq^2(q^2_0+q^2\mathrm{cos}^2\theta)}{(q^2-m^2_1) ((P-q)^2-m^2_2) ((k+q-P)^2-M^2_l)}\ .\label{eq:f1}
\end{eqnarray}
In the above equation we have performed the integral over the $\phi$ angle and $q_3=q\,\mathrm{cos}\,\theta$, being $\theta$ the angle from the $z$ axes. As can be observed, coefficients $A_1$ and $E_1$ contain logarithmically divergent integrals. To evaluate them, we perform the integral in $q_0$, and then the integral over the three-momentum using a cutoff, $q_{\mathrm{max}}$. The choice of $q_{\mathrm{max}}$ must be consistent with the maximum momenta needed in the two-meson loop integrals of the coupled-channel calculation, since here we have the same vertex of the resonance coupled to $D^*D^*_{(s)}$, as depicted in Fig. (\ref{fig:2}). In \cite{exotic}, the two-meson loop function is evaluated by means of the dimensional regularization, with $\mu=1500$ MeV and $\alpha=-1.4$. We redo the coupled channel calculation in \cite{exotic} using the formula for the two-meson loop function with cutoff, $q_{\mathrm{max}}$, which is,
\begin{equation}
G(s)=\frac{1}{4\pi^2} \int_0^{q_{\mathrm{max}}}q^2 dq
 \frac{ \omega_1 + \omega_2}{\omega_1\omega_2(P_0^2 - (\omega_1 + \omega_2)^2 + i \eps)}\ ,\label{eq:gcutoff}
\end{equation}
Where $\omega_{1,2}=\sqrt{q^2+m^2_{1,2}}$, and $m_{1,2}=m_{D^*}$ or $m_{D^*_{s}}$. We obtain the value of the cutoff needed to reproduce the states $R_{cc}$ and $S_{cc}$ in \cite{exotic}, $q_{\mathrm{max}}=750$ MeV, and couplings slightly higher than in \cite{exotic}: $g_R=20895$ MeV and $14705$ MeV for the $R_{cc}$ and $S_{cc}$ respectively. 

In Eq. (\ref{eq:int3}), the term not proportional to $q_\alpha q_\mu$ is convergent and can be evaluated by means of a Feynman parametrization.  The explicit form of the integrals in Eqs. (\ref{eq:f}) and (\ref{eq:f1}) are given in the Appendix (A.2). Finally, the amplitude of the diagram in Fig. \ref{fig:4}. 1), can be written as
\begin{eqnarray}
&&t^{ij}=I\,g_XgAG\rq{}V_3 \,\eps_\delta\lbrace H P_\alpha k_\gamma(k^i\eps^{\alpha j\gamma \delta}-k^j\eps^{\alpha i\gamma \delta})
+F P_\gamma k_\alpha P^\delta \eps^{ij\gamma\alpha}-3e(P-k)_\gamma \eps^{ij\gamma\delta }\rbrace\ ,\nonumber\\\label{eq:v}
\end{eqnarray}
with $H=(b+d)/2$ and $F=A+C$. The expressions of the coefficients that remain in Eq. (\ref{eq:int3}) are given in the Appendix (A.2).%The coefficients of $B$ and $D$ are proportional to $k_\mu \eps^\mu$ that we take equal to zero using the Lorentz condition for the vector meson. The coefficients of $a$ and $c$ are proportial to $P_j$, and are also taken zero. 

\subsection{Decay mediated by a pseudoscalar meson, Fig. \ref{fig:4}. 2)}

Once we have evaluated the diagram in Fig. \ref{fig:4}. 1), the one in Fig. \ref{fig:4}. 2), with pseudoscalar exchange, is similarly obtained. From Eq. (\ref{lag3}), the amplitude for the PPV vertex is
\begin{equation}
t_{PPV}=g\,P_V \eps^\mu_1(2(P-k)-q)_\mu\ .\label{eq:ppv}
\end{equation}

%Working in the approximation of neglecting the three-momenta of the vector meson as compared to the mass, $|\vec{p}|/m_V\simeq 0$,  $\mu$ and $\beta$ are spatial index since $\eps^0\simeq 0$. 
Summing over polarizations of $V_1,V_2$,  $(\eps_{1\,i}\eps_{2\,j}-\eps_{1\,j}\eps_{2\,i})\eps_{1\,k}\eps_{2\,l}=\delta_{ik}\delta_{jl}-\delta_{jk}\delta_{il}$, for the amplitude of the diagram in Fig. \ref{fig:4}. 2), we obtain
\begin{eqnarray}
&&-it^{ij}=\nonumber\\&&\frac{1}{2}gg_{R} G\rq{}AP_V I  k_\gamma \eps_{(f)\,\delta}
\int\frac{d^4 q}{(2\pi)^4}\frac{(P-q)^\alpha (\eps^{\,j \gamma\delta}_{\alpha}(2(P-k)-q)_i-\eps^{\,i\gamma\delta}_{\alpha}(2(P-k)-q)_j)}{((k-P+q)^2-m^2_l+i\eps)(q^2-m^2_1)((P-q)^2-m^2_2)}\ .\nonumber\\\label{eq:2}
\end{eqnarray}
The integral in Eq. (\ref{eq:2}), without the antisymmetric tensor $\eps^{\,j \gamma\delta}_{\alpha}$ can be parametrized using Lorentz covariance as follows
\begin{eqnarray}
& &\int\frac{d^4 q}{(2\pi)^4} \frac{(P-q)^\alpha (2(P-k)-q)^i}{((k-P+q)^2-m^2_l+i\eps)(q^2-m^2_1+i\eps)((P-q)^2-m^2_2+i\eps)}\nonumber\\
& &=i\,(\tilde{A}P^\alpha P^i+\tilde{B} P^\alpha k^i+\tilde{C} k^\alpha k^i+\tilde{D} k^\alpha P^i+\tilde{E} g^{\alpha i})
\end{eqnarray}
The terms with the coefficients $\tilde{A},\tilde{D}$ are proportional to the momenta of the initial particle, $P^i=0$, and the term that goes with $\tilde{C}$ is zero since the presence of $k_\alpha$ makes $k_\alpha k_\gamma\eps^{\alpha i \gamma \delta}=0$. Therefore, only $\tilde{B}$ and $\tilde{E}$ remain, and we obtain,
\begin{eqnarray}
t^{ij}=\frac{1}{2}g_{R}(G\rq{}g)AP_VI k_\gamma\eps_{(f)\,\delta}\lbrace \tilde{B} P_\alpha (\eps^{\alpha i\gamma\delta}k^j-\eps^{\alpha j\gamma\delta}k^i)-2\tilde{E}\eps^{ij\gamma\delta}\rbrace\ .\label{eq:p}
\end{eqnarray}
With $\tilde{B}$ and $\tilde{E}$ given in the Appendix (A.3).
\subsection{Total amplitude and decay width}
The sum of the amplitudes of the diagrams depicted in Fig. \ref{fig:4}, including pseudoscalar and vector meson exchange, Eqs. (\ref{eq:v}) and  (\ref{eq:p}), can be written as: 
\begin{eqnarray}
&&t_1^{ij}= g_{R}(G\rq{}g)\eps_{(f)\,\delta}\lbrace {\cal H}P_\alpha k_\gamma(k^i \eps^{\alpha j \gamma \delta}-k^j\eps^{\alpha i\gamma \delta})+ ( {\cal I}k_\gamma +{\cal J}P_\gamma )\eps^{ij\gamma\delta}+ {\cal F} P_\gamma k_\alpha P^\delta\eps^{ij\gamma\alpha}\rbrace \nonumber\\\label{eq:squ}
\end{eqnarray}
with ${\cal H}=(\sum_{\mathrm{diag}}(C_V H -\frac{1}{2}C_P \tilde{B}))$, ${\cal I}=(\sum_{\mathrm{diag}}(3e(M_l)C_V -e(m_l)C_P))$, ${\cal J}=-\sum_{\mathrm{diag}} 3e(M_l)C_V$, ${\cal F}=(\sum_{\mathrm{diag}} C_V F )$, and $C_V=AIV_3$, $C_P=AIP_V$. The squared amplitud, $\sum_{\delta,i,j}\vert t^{ij}_1\vert ^2$, is

\begin{eqnarray}
\sum_{\delta}\vert t_1\vert ^2=g^2 g_{X}^2G\rq{}^2({\it r_1} |\vec{k}|^4+ {\it r_2} |\vec{k}|^2+{\it r_3})\label{eq:squ1}\end{eqnarray}with\begin{eqnarray}
&&r_1=(2 P_0^4 |{\cal F}|^2)/m_{V_f}^2 + 4P_0^2 |{\cal H}|^2\ ,\nonumber\\
&&r_2=4 |{\cal I}|^2 + (2 P_0^2 |{\cal J}|^2)/m_{V_f}^2 + 
 4 P_0^2 Re({\cal I}{\cal F}^*) + (
 4 k_0 P_0^3 Re({\cal J}{\cal F}^*))/m_{V_f}^2 \ ,\nonumber\\&&+ 
 8 k_0 P_0 Re({\cal H}{\cal I}^*) + 8 P_0^2 Re({\cal H}{\cal J}^*)\ ,\nonumber\\
&& r_3=6 m_{V_f}^2 |{\cal I}|^2 + 6 P_0^2|{\cal J}|^2 + 
 12 k_0 P_0 Re({\cal I}{\cal J}^*)\ ,
\end{eqnarray}
%\begin{eqnarray}
%\sum_{\delta(\gamma)}\vert t\vert ^2=&&|{\cal H}|^2 (4 |\vec{k}|^4 P^2_0)+|{\cal I}|^2(6 k^2_0-2|\vec{k}|^2)+2 |{\cal F}|^2 P^4_0\frac{|\vec{k}|^4}{m^2_{D^*}}\nonumber\\&&+6 |{\cal J}|^2P^2_0+2|{\cal J}|^2 P^2_0\frac{|\vec{k}|^2}{m^2_{D^*}}+8\mathrm{Re} {\cal H}{\cal I}^*P^0k^0|\vec{k}|^2\nonumber\\&&+4\mathrm{Re} {\cal J}{\cal F}^*P^3_0k_0\frac{|\vec{k}|^2}{m^2_{D^*}}-2\mathrm{Re}\left( -4 {\cal H}{\cal J}^* P^2_0|\vec{k}|^2-6{\cal I}{\cal J}^*k^0P^0-2{\cal I}{\cal F}^*P^2_0|\vec{k}|^2\right)
%\end{eqnarray}
where $m_{V_f}=m_{D^*}$. In practise, there is only one constant which is complex, ${\cal I}$ for the diagram with $m_l=m_\pi$. The coefficients $C_V$ and $C_P$ needed in Eqs. (\ref{eq:squ}) and (\ref{eq:squ1}) are giving in Tables \ref{tab:coef1} - \ref{tab:coef6}. % So, we can write,
%\begin{eqnarray}
%&&\sum_{\delta}\vert t\vert ^2=g^2 g_{X}^2G\rq{}^2({\cal A}(|\vec{k}|^2+{\cal B})^2+{\cal C})\,\,\mathrm{with}:\nonumber\\
%&&{\cal A}=4 {\cal H}^2 P_0^2 + \frac{(2{\cal F}^2 P_0^4)}{m_{V_f}^2}\nonumber\\
%&&{\cal B}=\frac{2 m_{V_f}^2 |{\cal I}|^2 + 
  % P_0 ({\cal J} P_0 ( {\cal J} + 4 {\cal H} m_{V_f}^2 + 2 {\cal F} k_0 P_0) + 
  %    m_{V_f}^2 (4 {\cal H} k_0 + 2 {\cal F} P_0) Re({\cal I}))}{4 {\cal H}^2 m_{V_f}^2 P_0^2 + 
  % 2 {\cal F}^2 P_0^4}\nonumber\\
%&& {\cal C}=6 {\cal J}^2 P_0^2 + 6 m_{V_f}^2 |{\cal I}|^2 + 
 %12 {\cal J} k_0 P_0 Re(
 %  {\cal I}) \nonumber\\&&- \frac{ (2 m_{V_f}^2 |{\cal I}|^2 + 
   %   P_0 ({\cal J} P_0 ({\cal J} + 4 {\cal H} m_{V_f}^2 + 2 {\cal F} k_0 P_0) + 
   %      m_{V_f}^2 (4 {\cal H} k_0 + 2 {\cal F} P_0) Re({\cal I})))^2}{4 {\cal H}^2 m_{V_f}^4 P_0^2 + 
   % 2 {\cal F}^2 m_{V_f}^2 P_0^4}\ .
%\end{eqnarray}
Finally, the decay width is given by:
\begin{equation}
\Gamma_{R_{cc}\to DD^*_{(s)}}=\frac{1}{2J+1}\frac{|\vec{k}|\sum_{\delta}\vert t_1\vert ^2}{8\pi M_{R_{cc}}^2}
\end{equation}

In Eq. (\ref{eq:ppv}),  to reproduce the experimental decay width $D^{*+}\to D^0\pi^+$, with $P_V=-1$, a value $g=g_D=8.95$ is needed, which is significantly larger than that expected from 
SU(4), g = 4.16.  However  ($G^\prime g$), which is the factor that appears in the amplitude of the process $R_{cc}\to DD^*$,  see Eq. (\ref{eq:squ}), is not so different from what we expect from SU(4), because at the same time $\Gamma^{\mathrm{exp}}_{D^{*+}\to D^+\gamma}$ is smaller than what one obtains from the hidden gauge Lagrangian, and this can be compensated taking a value of $G\rq{}$ smaller ($G\rq{}=0.0087$ MeV$^{-1}$ see section 5). %the $G^\prime$ in the charm sector is smaller than the one of the light sector.  %is needed. However, the final amplitude $t$ for the process $R_{cc}\to DD^*_{(s)}$ is proportional to the product $G\rq{} g$, which is not so different from SU(3) to SU(4) (see Section 5). 
Thus, we use the value of $(G\rq{} g)=0.06$ MeV$^{-1}$, obtained with $M_V=m_\rho$ and $f=f_\pi=93$ MeV, $g=M_V/2f$, and  when performing the evaluation of the errors, we vary the ($G\rq{}g$) constant from this value. Also, one must take into account that using this value of $g$, $g_l$ or $g_h$, the subtraction constant $\alpha_h$ in \cite{exotic} can be tunned to obtain the properly masses as explained in section 2. 
\section{Radiative decays of doubly charmed molecules into $DD\gamma$}

\begin{figure}
\begin{center}
\includegraphics[width=16cm]{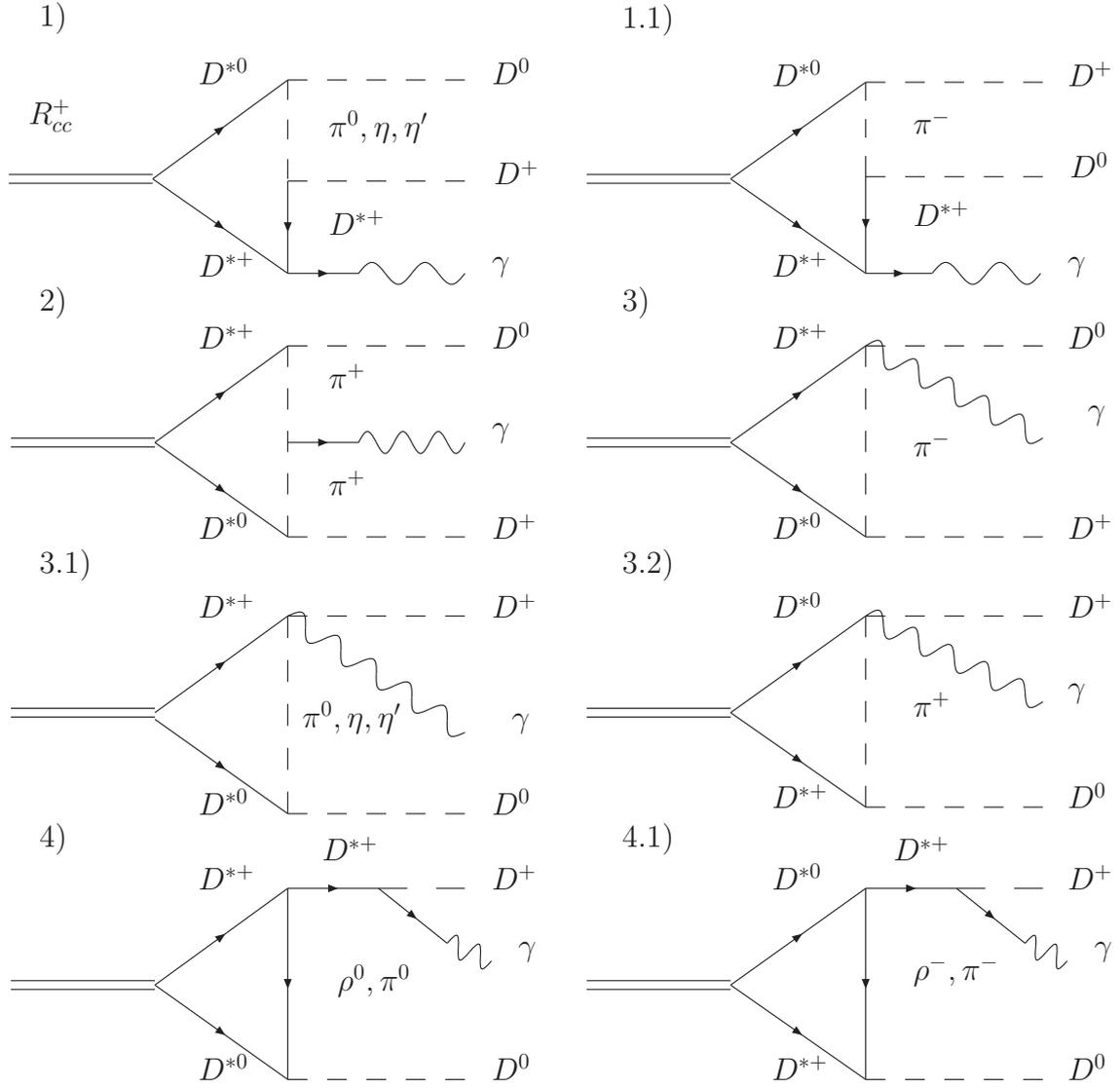}
\end{center}
\caption{ diagrams for the $R_{cc}^+\to D^0D^+\gamma$ decay through one loop.}
\label{fig:radocharm11} 
\end{figure}

In this section we consider the radiative decay of doubly charmed meson molecules,  $R_{cc}\to DD\gamma$. In our picture, the decays occur via the one loop diagrams depicted in Fig. \ref{fig:radocharm11}.
Only non-vanishing diagrams are shown: the contact vertex $D^{*0}D^0\gamma\pi^0$ leads to a zero amplitude (also, when the photon comes out from a neutral vector or pseudoscalar meson, the sum for all the intermediate neutral particles $\rho$, $\omega$, $J/\psi$, gives zero.  In the last two diagrams, 4) and 4.1), it is also possible to have the vertex $D^+D^+\gamma$ instead of $D^{*+}D^+\gamma$. However, since the initial state has $J^P=1^+$, the intermediate transition $D^*D^*\to DD$ is not possible. Below, we evaluate the different structures 1), 2) and 3) depicted in Fig. \ref{fig:radocharm11}. The decay width coming from diagrams 4) and 4.1) can be easily obtained from the evaluation of the decay width to $D_{(s)}D^*_{(s)}$ done in section 3, and their contributions is discussed in the Results section (6).
\begin{figure}
\begin{center}
\includegraphics[width=8cm]{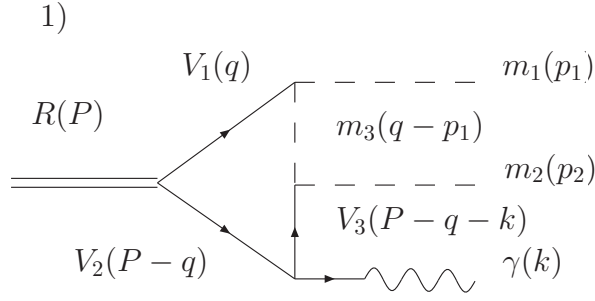}
\end{center}
\caption{Momentum dependence of the first diagram in Fig. \ref{fig:radocharm11}.}
\label{fig:radocharm31} 
\end{figure}
\subsection{Evaluation of the diagram in Fig. \ref{fig:radocharm11} 1).}
To evaluate the diagram in Fig. \ref{fig:radocharm11} 1), we show explicitly the momentum assignment in Fig. \ref{fig:radocharm31}. The necessary interaction Lagrangians are given in Eq.  (\ref{lag3}). In addition, the hidden gauge Lagrangian, \cite{hideko}, provides the coupling of the vector meson dominance,
\begin{equation}
t_{V\gamma}={\cal P} M^2_V\frac{e}{g}\eps_{(\gamma)\alpha}\eps^\alpha_{V}\ .
\end{equation}
Essentially, it replaces $\eps(V)$ by $\eps(\gamma)$ in Eq. (\ref{eq:3vertex}). Then, we can write the amplitude as
\begin{eqnarray}
&&-it_{ij}^{1)}=\frac{1}{2}{\cal A}g^2 \tilde{g}  e\eps_{(\gamma)\beta}\int\frac{d^4 q}{(2\pi)^4}(\eps_{1\,i}\eps_{2\,j}-\eps_{1\,j}\eps_{2\,i})\eps^\mu_1\eps^{\mu\rq{}}_3
 (2p_1-q)_\mu (p_1-p_2-q)_{\mu\rq{}}\nonumber\\&&\times\lbrace (2k+q-P)_\alpha \eps_{3}^\beta\eps^\alpha_2-(k+P-q)_\alpha\eps_{2}^{\beta}\eps^\alpha_3+(2(P-q)-k)^\beta\eps_{\alpha\,3}\eps^\alpha_2\rbrace\nonumber\\&&\times \frac{1}{(q-p_1)^2-m^2_3 }\frac{1}{q^2-M^2_1}\frac{1}{(P-q)^2-M^2_2}\frac{1}{(P-q-k)^2-M^2_3}\ .\nonumber\\\label{eq:tddp}
\end{eqnarray}
with ${\cal A}=I  V_3 {\cal P}P_V$. The sum over polarizations of vectors 1, 2, 3, gives,
\begin{eqnarray}
&&\sum \eps^i_1\eps^\mu_1\simeq-g^{i\mu}\nonumber\\
&&\sum \eps^j_2\eps^\alpha_2\simeq-g^{j\alpha}\nonumber\\
&&\sum \eps^\beta_3\eps^{\mu\rq{}}_3=-g^{\beta\mu\rq{}}+\frac{(P-q-k)^{\mu\rq{}}(P-q-k)^\beta}{m^2_3}\ .\label{eq:int}
\end{eqnarray}
In the last sum in Eq. (\ref{eq:int}), it is possible to make an approximation without loosing much in the numerical results. For the $D^*D^*$-molecule decaying into $DD\gamma$, the intermediate $D^*$ of momentum $P-q-k$ is also close to mass-shell. In fact, when the photon takes its maximum energy, $E_{\gamma}^{\mathrm{max}}=(s-4 m_{D}^2)/2 \sqrt{s}$, we have, for $\beta=\mu=0$,  $-1+\frac{(P-q-k)^{0}(P-q-k)^0}{m^2_{D^*}}\simeq0.23$ and for $\beta=0$, $\mu=i$, $\frac{(P-q-k)^{0}(P-q-k)^i}{m^2_{D^*}}\simeq0.1$. We make the approximation of neglecting the zero component of the last sum over polarizations in Eq. (\ref{eq:int}), which simplifies considerably the calculation. In this framework, the amplitude in Eq. (\ref{eq:tddp}) can be written as:
\begin{eqnarray}
&&-it_{ij}^{1)}=\frac{1}{2}{\cal A}eg^2 \tilde{g}\eps_{(\gamma)m }\int\frac{d^4q}{(2\pi)^4}\frac{1}{(q-p_1)^2-m^2_3}\frac{1}{q^2-M^2_1}\frac{1}{(P-q)^2-M^2_2}\nonumber\\&&\times \frac{1}{(P-q-k)^2-M^2_3} \lbrace(2p_1-q)_i((2k+q)_j(2p_1+k-q)_{m}\nonumber\\&&-(2p_1+k-q)_{l}(k-q)_{l}\delta_{jm}+(-2q-k)_m(2p_1+k-q)_j)\nonumber\\&&-(\mathrm{same \,but} \, i\leftrightarrow j)\rbrace\ .\nonumber\\\label{eq:t1}
\end{eqnarray}
 The result of this integral, (without the factor $\frac{1}{2}{\cal A}eg^2 \tilde{g}\eps_{(\gamma)}^m$) takes the form:
\begin{eqnarray}
&&i(a_1(k_i\delta_{jm}- k_j\delta_{im}) +c_1(k_j p_{1\,i} p_{1\,m}-k_i p_{1\,j} p_{1\,m})+d_1(p_{1\,i}\delta_{jm} -p_{1\,j}\delta_{im} ))\ .\nonumber\\\label{eq:lor}
\end{eqnarray}
In the above expression, we have omitted the term $e_1(p^i_1k^jk^m-p^j_1 k^i k^m)$, which disappear when contracting with $\eps_{(\gamma)}^m$ in the Coulomb gauge . %Whereas, for the second term in Eq. (\ref{eq:t1}), we have,
%\begin{eqnarray}
%i\lbrace a_1(g^{i\beta}k^j-g^{\beta j} k^i) +e_1(g^{\beta j} p^i_1-g^{i\beta} p^j_1)\rbrace\ .\label{eq:lor1}
%\end{eqnarray}
The integral in Eq. (\ref{eq:t1}) is convergent and we use the formula for the Feynman parametrization for $n=4$ to evaluate it,
\begin{eqnarray}
\frac{1}{abcd}=6\int^1_0 dx\int^x_0 dy \int^y_0 \frac{dz}{\left( a+(b-a) x+(c-b) y+(d-c) z\right)^4}\label{eq:parfe}
\end{eqnarray}
with 
\begin{eqnarray}
&&a=q^2-M^2_1\nonumber\\
&&b=(q-p_1)^2-m^2_3\nonumber\\
&&c=(P-q)^2-M^2_2\nonumber\\
&&d=(P-q-k)^2-M^2_3\ .
\end{eqnarray}
The change of variable, $q=q\rq{}+p_1 x+(P-p_1) y-kz$, simplifies the denominator in Eq. (\ref{eq:parfe}) as $(q\rq{}^2+s)^4$ with
%\begin{eqnarray}
%&&\int\frac{d^4 q\rq{}}{(2\pi)^4} 6\int^1_0 dx\int^x_0 dy \int^y_0\frac{dz}{(q\rq{}^2+s)^4}(2p_1-q)^i \lbrace(2k+q-P)^j(2p_1-P-q+k)^\beta\eps_\beta(\gamma)\nonumber\\&&-\eps^j(\gamma)(2p_1-P+k-q)^\alpha (k+P-q)_\alpha+\eps^\beta(\gamma)(2(P-q)-k)_\beta(2p_1-P+k-q)^j\rbrace\nonumber\\&&-(\mathrm{same\,but\, i\leftrightarrow j})\ .
%\end{eqnarray}
%with $q=q\rq{}+p_1 x+(P-p_1) y-kz$ and 
\begin{eqnarray}
&&s=(p^2_1+M^2_1-m^2_3) x+(P^2-M^2_2-p^2_1+m^2_3)y+(-2Pk-M^2_3+M^2_2) z\nonumber\\&&-M^2_1-(p_1 x+(P-p_1) y -kz)^2\ .
\end{eqnarray}
Picking up the coefficients of $ k_i\delta_{m j}$, $k_jp_{1\,i}p_{1\,\beta}$, $p_{1\,i}\delta_{m j} $, keeping the antisymmetric combination in the indices $i,j$, we obtain,
\begin{eqnarray}
&&t_{ij}^{1)}=-\frac{1}{2}{\cal A}eg^2 \tilde{g}\eps_{(\gamma) m}(a_1(k_i\delta_{jm}- k_j\delta_{im}) +c_1(k_j p_{1\,i} p_{1\,m}-k_i p_{1\,j} p_{1\,m})+d_1(p_{1\,i}\delta_{jm} -p_{1\,j}\delta_{im} ))\ ,\nonumber\\\label{eq:ta}
\end{eqnarray}
And the coefficients $a_1$, $c_1$ and $d_1$ are given in the Appendix (A.4).
%\begin{eqnarray}
%\int\frac{d^4 q}{(q^2+s)^4}=\frac{i\pi^2}{6s^2}\nonumber\\
%\int\frac{d^4 q q^2}{(q^2+s)^4}=\frac{i\pi^2}{3s}\nonumber\\
%\int\frac{d^4 q \vec{q}\,^2}{(q^2+s)^4}=-\frac{i\pi^2}{4s}\nonumber\\
%\end{eqnarray}
%\begin{eqnarray}
%&&\sum_\delta|t|^2={\cal C}eg^2 \tilde{g}\frac{1}{|\vec{k}|^2}\left(2 (2 A^2 |\vec{k}|^4 + C^2 (\vec{k}\cdot \vec{p}_1\,^2- |\vec{k}|^2 |\vec{p}_1|^2)^2 + 
  % 2 C D \vec{k}\cdot \vec{p}_1 (\vec{k}\cdot \vec{p}_1\,^2 \right .\nonumber\\&&\left .- |\vec{k}|^2 |\vec{p}_1|^2) + D^2 (\vec{k}\cdot \vec{p}_1\,^2 + |\vec{k}|^2 |\vec{p}_1|^2) + 
  % 2 A |\vec{k}|^2 (2 D \vec{k}\cdot \vec{p}_1+ c (\vec{k}\cdot \vec{p}_1\,^2 - |\vec{k}|^2 |\vec{p}_1|^2)))\right)
   %\end{eqnarray}
 %\begin{eqnarray}
 %4 A^2 \vec{k}\,^2 + 8 AD \vec{k}\cdot\vec{ p }_1+ 4 A C (\vec{k}\cdot\vec{ p }_1)^2 + 4 D^2 \vec{p}_{1}\,^2 - 
 %4 A C \vec{k}\,^2 \vec{p}_1\,^2 - 2 C^2 (\vec{k}\cdot\vec{ p }_1)^2 \vec{p}_1\,^2 + 2 C^2 \vec{k}^2 \vec{p}_1\,^4
 %\end{eqnarray}
\subsection{Evaluation of diagrams Fig. \ref{fig:radocharm11} 2) and 3).}
\begin{figure}
\begin{center}
\includegraphics[width=13cm]{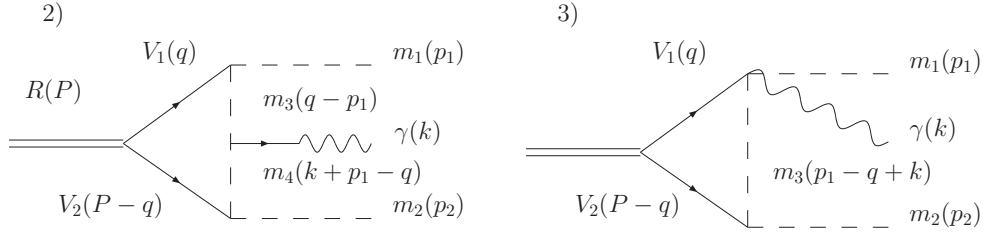}
\end{center}
\caption{Momentum dependence of the Feynman diagrams type 2) and 3) in Fig. \ref{fig:radocharm11}.}
\label{fig:radocharm41} 
\end{figure}
Let us consider the relevant diagrams in Fig. \ref{fig:radocharm41}, with momentum variables shown explicitly. For diagram 2), the amplitude is:
\begin{eqnarray}
&&-it_{ij}^{2)}=\frac{1}{2}{\cal B} eg^2 \tilde{g} \eps_{(\gamma)}^\alpha\int \frac{d^4q}{(2\pi)^4}\frac{1}{q^2-M^2_1}\frac{1}{(q-p_1)^2-m^2_3}\frac{1}{(p_1+k-q)^2-m^2_4}\frac{1}{(P-q)^2-M^2_2}\nonumber\\&&\times(2(q-p_1)-k)_\alpha\lbrace(2p_1-q)_i(p_1-p_2+k-q)_j-(2p_1-q)_j(p_1-p_2+k-q)_i\rbrace\nonumber\\\label{eq:diag2}
\end{eqnarray}
with ${\cal B}=  I {\cal P} P_V$, and $P_V=P_{V_1}P_{V_2}P_{V_3}$ (the product of the coefficients for the $PPV$ vertices in Eq. (\ref{eq:ppv})). The integral in Eq. (\ref{eq:diag2}) must be like:
\begin{equation}
i(a_2(k_ig_{\alpha j}-k_jg_{\alpha i})+b_2P_\alpha(p_{1\,i}k_j-k_ip_{1\,j})+c_2 ( k_jp_{1\,i}-k_ip_{1\,j})p_{1\,\alpha}+d_2(p_{1\,i}g_{\alpha j}-p_{1\,j}g_{\alpha i}))\ .
\end{equation}
We apply the change of variable $q=q\rq{}+(P-p_1)y+(p_1-P+k)z $ with the same Feynman parametrization of Eq. (\ref{eq:parfe}), but now $d=(p_1-q+k)^2-m^2_4$. In the Coulomb gauge for the polarization vector of the photon, we can take $\eps^0_{(\gamma)}=0$ and $\eps^i_{(\gamma)}k_i=0$, we can neglect the second term. We obtain,
\begin{eqnarray}
&&t_{ij}^{2)}=-\frac{1}{2}{\cal B} eg^2 \tilde{g} \eps_{(\gamma)\,m}(a_2(k_i\delta_{m j}-k_j\delta_{m i})-c_2 ( k_jp_{1\,i}-k_ip_{1\,j})p_{1\,m}+d_2(p_{1\,i}\delta_{m j}-p_{1\,j}\delta_{m i}))\ ,\nonumber\\\label{eq:tb}\end{eqnarray}
which has the same form than Eq. (\ref{eq:ta}). The coefficients $a_2$, $c_2$ and $d_2$ are given in the Appendix (A.4). 
   
     To evaluate the diagram in Fig. \ref{fig:radocharm11} 3), we use the Lagrangian \cite{hideko}
    \begin{eqnarray}
    {\cal L}_{V\gamma PP}=\frac{e M^2_V}{4g f^2}A_\mu \langle V^\mu(Q\Phi^2+\Phi^2 Q-2\Phi Q\Phi)\rangle\ ,
    \end{eqnarray}
  for the $V\gamma PP$ vertex, with $Q=\frac{1}{3}diag(2,-1,-1,2)$, which leads to 
\begin{equation}
t_{V\gamma PP}= V_{2P\gamma} \eps^\nu_1\eps_{(\gamma)\,\nu}\ .
\end{equation}
Thus the amplitude of diagram Fig. \ref{fig:radocharm41} 3) is
    \begin{eqnarray}
   t_{ij}^{3)}=-\frac{1}{2}{\cal C}eg^2\tilde{g} \eps_{(\gamma)m}a_3( (k_i\delta_{m j}-k_j \delta_{m i})+(p_{1\,i}\delta_{m j}-p_{1\,j} \delta_{m i}))\label{eq:tc}
    \end{eqnarray}
   where ${\cal C}=IP_{V_1}V_{2P\gamma}$ and $a_3$ is given in the Appendix (A.4).
\subsection{Total amplitude}
So far we have seen that the amplitude takes the structure of Eq. (\ref{eq:ta}) with the coefficients contributed from various diagrams. Therefore, the square amplitude takes the following form with the coefficients computed from Eqs. (\ref{eq:ta}), (\ref{eq:tb}) and (\ref{eq:tc})
%Finally, the square of Eq. (\ref{eq:ta}) taking into account all the diagrams in Fig. \ref{fig:radocharm11} and all the different structures in Eqs.   is 
\begin{eqnarray}
&&\sum_{\delta\,ij}|t_{2\,ij}|^2=\frac{1}{4}\frac{eg^2\tilde{g} }{|\vec{k}|^2}(2 (C^2 f(\vec{k},\vec{p}_1)^2 + 2 A C f(\vec{k},\vec{p}_1) \vec{k}\,^2 + 2 A^2 \vec{k}\,^4 +   2 D \vec{k}\cdot \vec{p}_1 (C f(\vec{k},\vec{p}_1) + 2 A \vec{k}\,^2) \nonumber\\&&
 + D^2 (f(\vec{k},\vec{p}_1) + 2 \vec{k}^2 \vec{p}_1\,^2)))\label{eq:t2f}\end{eqnarray}
 with
 \begin{eqnarray}
  f(\vec{k},\vec{p}_1) = (\vec{k}\cdot \vec{p}_1)\,^2 - |\vec{k}|^2 |\vec{p}_1|^2
   \end{eqnarray}
   and
   \begin{eqnarray}
   &&A={\cal A}a_1+{\cal B}a_2+{\cal C}a_3\nonumber\\
   &&C={\cal A}c_1-{\cal B}c_2\nonumber\\
   &&D={\cal A}d_1+{\cal B}d_2+{\cal C}a_3\ .\label{eq:coefs}
     \end{eqnarray}
  In addition, if we call $p_1$ to the momenta of the $D^0$ in the final state, in order to consider all diagrams in Fig.  \ref{fig:radocharm11} including those where it is placed $D^+$ with momenta $p_2$ instead of $D^0$ (diagrams of Fig.  \ref{fig:radocharm11} 1.1), 3.1) and 3.2)) we do in Eq. (\ref{eq:t2f}),
   \begin{eqnarray}
  && A\to A+A\rq{}-D\rq{}\nonumber\\
  && C\to C+C\rq{}\nonumber\\
  && D\to D-D\rq{}\ ,\label{eq:coefs2}
   \end{eqnarray}
   where $A$, $C$ and $D$ of the r. h. s of Eq. (\ref{eq:coefs2}) are those coefficients in Eq. (\ref{eq:coefs}) and $A\rq{}$, $C\rq{}$, $D\rq{}$, the same but changing $p_1\to p_2=P-p_1-k$ and $p_2\to p_1$.
   
   The final decay width of the process $R_{cc}^+(P)\to D^0(p_1)D^+(p_2)\gamma(k)$ for the initial particle at rest, is
   \begin{eqnarray}
   &&\Gamma=\frac{1}{64M_X\pi^3(2J+1)}\int_{E_{1\,\mathrm{min}}}^{E_{1\,\mathrm{max}}}dE_1\int_{E_{\gamma \,\mathrm{min}}}^{E_{\gamma \,\mathrm{max}}}dE_\gamma\,\theta(1-\mathrm{cos}^2\theta )\sum_{\delta}|t_2|^2 \,\,\ ,\mathrm{with}\nonumber\\\nonumber\\
   &&\mathrm{cos}\,\theta =\frac{(M_X-E_1-E_\gamma)^2-m^2_2-|\vec{p}_1|^2-|\vec{k}|^2}{2|\vec{p}_1||\vec{k}|}\nonumber\\
   &&E_{1\,\mathrm{min}}=m_1\,\ ,E_{2\,\mathrm{max}}=\frac{s+m^2_1-m^2_2}{2\sqrt{s}}\nonumber\\
   &&E_{\gamma\,\mathrm{min}}=0\, \ ,E_{\gamma\,\mathrm{max}}=\frac{s-m^2_1-m^2_2-2m_1m_2}{2\sqrt{s}}\ ,
   \end{eqnarray}
   being $\theta$ the angle between $p_1$ and $k$.

\section{Decay of doubly charmed state to $D^*D\gamma$}
\begin{figure}
\begin{center}
\includegraphics[width=10cm]{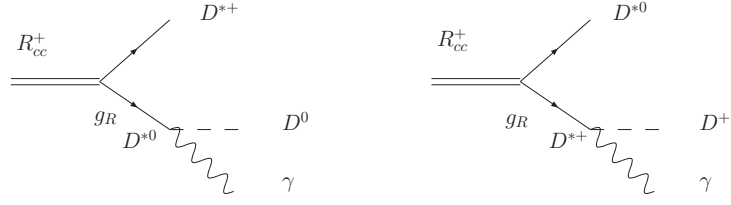}
\end{center}
\caption{Decay of the doubly charmed states into $D^*D_{(s)}\gamma$}
\label{fig:doddsg} 
\end{figure}

Finally, we consider the radiative decay depicted in Fig. \ref{fig:doddsg}. In order to evaluate both diagrams for $R_{cc}^+\to D^{*0}D^+\gamma$ and $R_{cc}^+\to D^{*+}D^0\gamma$, we need the amplitude for the decay $D^{*+}\to D^+\gamma$ and $D^{*0}\to D^0\gamma$. They are given by the Lagrangian in Eq. (\ref{lagan}), and the amplitude is 
\begin{eqnarray}
&&t_{(D^*\to D\gamma)}=-\frac{G\rq{}}{\sqrt{2}}\frac{e}{g}C\rq{} q_\mu \eps_{(D^{*+})\,\nu} k_\alpha\eps_{(\gamma)}\eps^{\mu \nu\alpha\beta}\,\,\mathrm{with} \,\,C\rq{}=\left \{\begin{array}{c}\,\,\frac{4}{3}\,\, \mathrm{for}\,\, D^{*0}\\\\\frac{1}{3}\,\,\mathrm{for}\,\,D^{*+}\\\\\frac{1}{3}\,\,\mathrm{for}\,\,D^{*+}_s\end{array}\right .
\end{eqnarray}
Fixing the coupling $g=m_\rho/ 2 f_\pi=4.16$, we can use a value of $G\rq{}$ in order to reproduce the experimental decay widths from the PDG, which is $\Gamma(D^{*+}\to D^+\gamma)/\Gamma_{D^{*+}}=(1.6\pm 0.4)\%$, with $\Gamma_{D^{*+}}=96\pm 22$ KeV. We obtain $G\rq{}=0.0087$ MeV$^{-1}$.

  Now the amplitude of the processes depicted in Fig. \ref{fig:doddsg},
\begin{eqnarray}
&&t_{3\,ij}=I\, g_R\frac{1}{2}\frac{G\rq{}}{\sqrt{2}}\frac{e}{g}C\rq{}\frac{1}{q^2-m^2_{D^*}}q_\mu k_\alpha \eps_{(\gamma)\,\beta}\left( \eps^{\mu\,\,\alpha \beta}_{\,\,j}\eps_{1\,i}-\eps^{\mu\,\, \alpha\beta}_{\,\, i}\eps_{1\,j}\right)\ .
\end{eqnarray}
Taking the square of this, we find the decay width
\begin{eqnarray}
&&\Gamma=\frac{1}{32 \,M_R \pi^3}\int \frac{\tilde{p}_2p_1}{\sqrt{s}}\frac{1}{2 J+1}\sum_\delta \vert t_3\vert^2 d M_{D\gamma}\\
\end{eqnarray}with\begin{eqnarray}
&&\sum_\delta \vert t_3\vert^2=-\frac{C\rq{}^2I^2 g_R^2G\rq{}^2 e^2}{4 g^2}\left(\frac{M_{D\gamma}^2-m^2_D}{M^2_{D\gamma}-m^2_{D^*}}\right)^2\end{eqnarray}and \begin{eqnarray}
&&\mathrm{and}\,\,\tilde{p}_2=\frac{\lambda^{1/2}(M_{D\gamma}^2,0,m_D^2)}{2\, M_{D\gamma}}\ ,p_1=\frac{\lambda^{1/2}(M_R^2,m_{D}^2,M_{D\gamma}^2)}{2\, M_{R}}\nonumber .
\end{eqnarray}
where $M_{D\gamma}$ is the invariant mass of the $D$ meson and $\gamma$.
\section{Results}

In this section we discuss the numerical results of the decays evaluated in the previous sections.% In order to be consistent with the use of the cutoff,  $q_{\mathrm{max}}$, in the formulas Eqs. (\ref{eq:e}), (\ref{eq:a1}), (\ref{coff1}) and (\ref{coff2}), we redo the coupled channel calculation of \cite{exotic} using the formula of Eq. (\ref{eq:gcutoff}). 
   They are shown in Table \ref{tab:res1}.  Here, $\Gamma_{\mathrm{tot}}$ is the full width, $\Gamma^k$ are the partial decay widths for the two possible channels $k=1,2$, ex. $D^0D^{*+}$ and $D^+D^{*0}$, and $\Gamma^k_j$ is $\Gamma_{R(S)\to D_{(s)}D\pi}=\Gamma^k \Gamma_{D^*\to D\pi}/\Gamma_{D^*}$ ({\it{hadronic decays}}) or $\Gamma_{R(S)\to DD_{(s)}\gamma}=\Gamma^k\Gamma_{D^*_{(s)}\to D_{(s)}\gamma}/\Gamma_{D^*_{(s)}}$  ({\it{radiative decays})}. The partial decay widths of $D^*$ and $D^*_s$ to $D\pi$ and $D_{(s)}\gamma$, are taken from the PDG. The error shown is evaluated combining both, theoretical and the experimental errors from PDG. 
 
 To evaluate the errors we have considered variations of all parameters involved in the calculation. However some of them are not independent, and the evaluation of the error has to be done carefully. In the case of the decay into $DD^*_{(s)}$, we have the following parameters: ($g G\rq{}$), $q_{\mathrm{max}}$ and $g_R$. As mentioned, the coupling of the resonance, $g_R$ to the $D^*D^*_{(s)}$ and $q_{\mathrm{max}}$ are not independent. Furthermore, the mass of the states that we have found at $M_R= 3970$ MeV and $4100$ MeV also depends on $q_{\mathrm{max}}$: the bigger the cutoff is, the more bound the obtained state is, and the coupling $g_R$ also grows. Thus, the main source of uncertainties come from variations in the parameters ($g G\rq{}$) and $q_{\mathrm{max}}$. Performing the coupled channel calculation for several $q_{\mathrm{max}}$ around $750$ MeV, using linear regression we have found the following approximated relations valid close to the masses of the resonances $R_c(S_c)$, between the binding energy, $B$, defined as $B=\sqrt{s}_{th}-M_R$, $q_{\mathrm{max}}$ and $g_R$: 
\begin{eqnarray}
&&B_1=-110+0.21 \,q_{\mathrm{max}};\hspace{1cm} g_{1,\,R}=-929+28 \,q_{\mathrm{max}}\,\,\mathrm{for}\,\, R_{cc}(3970)\nonumber\\
&&B_2=-88+0.144 \,q_{\mathrm{max}};\hspace{1cm}g_{2,\,R}=-13216+37 \,q_{\mathrm{max}}\,\,\mathrm{for}\,\,S_{cc}(4100)\ ,
\end{eqnarray}
for $i=1,2$, the doubly charmed states without and with strangeness respectively. Taking a gaussian distribution around the mean values of ($g G\rq{}$) and $q_{\mathrm{max}}$ with $\sigma=0.15\mu$, and using the above relations, we obtain the errors shown in Table \ref{tab:res1}. 

 We observe that the total widths of the doubly charmed states, which comes basically from the decay into $D^0D^{*+}$ and $D^+D^{*0}$ for the $R_{cc}^+$, from the channels $D^0D^{*+}_s$ and $D^+_sD^{*0}$ for the $S_{cc}^+$,  $D^+_sD^{*+}$ and $D^+D^{*+}_s$ for the $S_{cc}^{++}$, are $(44\pm 12)$, $(24\pm 8)$, and $(24\pm 8)$ MeV for the $R^+_{cc}$, $S^+_{cc}$ and $S^{++}_{cc}$ respectively, giving both channels (ex. $D^0D^{*+}$ and $D^+D^{*0}$) the same contribution to the width. 
 
 The direct diagrams with three/four propagators of Fig. \ref{fig:radocharm11}, type $1)$, $2)$ and $3)$, lead to a very small width of the order of few KeV in the case of the $R_{cc}^+(3970)$ and $S_{cc}^+(4100)$ and $0.13$ KeV for the doubly charge state, $S_{cc}^{++}(4100)$, the difference in one order of magnitude is due to the lack of type $2)$ diagram in this case. The $D^*D_{(s)}\gamma$ decay channel is smaller than $1$ KeV, except for the channel $D^{*+}_sD^0\gamma$ channel of the $R_{cc}^+$ state, which is $4$ KeV. Direct diagrams with four propragators like in Fig. \ref{fig:radocharm11}, type $1)$, substituting the final vector line by a pseudoscalar, would also be possible for $DD_{(s)}\pi$ in the final state, however, they have anomalous couplings together with heavy meson propagator, for what we expect them to be also very small. 
 
 Thus, the diagrams responsible of the radiative decay of the doubly charmed molecules are those of type $4)$ in Fig. \ref{fig:radocharm11}, where the photon comes out from a $D^*$ meson. The width from these diagrams is obtained from section 3, multiplying $\Gamma^k$, the width of $R_{cc}\to DD^*$ by $\Gamma_{D^*\to D\gamma}/\Gamma_{D^*}$, the last quantity taken from the PDG. 
 
 Therefore, the most important decay modes of the doubly charmed mesons are $DD\pi$, $DD\gamma$, where $D\pi$ and $D\gamma$ comes from the decay of $D^*$, and $DD_s\pi$, $DD_s\gamma$, for the strangeness ones. 
 
 Finally, we perform the evaluation of the $R_{cc}\to DD^*$ decay width, shown in Fig. \ref{fig:3}, for only one meson exchanged, $\pi$, $\rho$, ... . The result is shown in Table \ref{tab:par1} for each doubly charmed meson. We can see that the most important contribution comes from $\rho$ exchange, then $\pi$ and $\omega$ respectively for the $R_{cc}^+(3970)$, where the interaction of the $\rho$ and $\pi$ are constructive, and opposite sign for the $\omega$. For the $S_{cc}(4100)^{+(+)}$, the most important contribution comes from the $K^*$, then, $K$ and $J/\psi$, being the $J/\psi$ exchange of opposite sign than the $K^*, K$. 
 
 %From Table \ref{tab:res1} we obtain the ratios,
%\begin{eqnarray}
%\frac{\Gamma(R_{cc}^+\to D^0 D^0 \pi^+)}{\Gamma(R_{cc}^+\to D^+D^0\pi^0)}=0.7,\hspace{1cm} \frac{\Gamma(R_{cc}^+\to D^+ D^0 \gamma)}{\Gamma(R_{cc}^+\to D^+D^0\pi^0)}=0.4
%\end{eqnarray}
%for the $R_{cc}^+(3970)$,
%\begin{eqnarray}
%\frac{\Gamma(S_{cc}^+\to D^+_sD^0 \pi^0)}{\Gamma(S_{cc}^+\to D^+_sD^0\gamma)}=0.44
%\end{eqnarray}
%for the $S_{cc}^{+}(4100)$, and 
%\begin{eqnarray}
%\frac{\Gamma(S_{cc}^{++}\to D^+_s D^+ \pi^0)}{\Gamma(S_{cc}^{++}\to D^+_sD^0\pi^+)}=0.5,\hspace{1cm} \frac{\Gamma(S_{cc}^{++}\to D^+ D^+_s \gamma)}{\Gamma(S^{++}_{cc}\to D^+_sD^0\pi^+)}=1.4
%\end{eqnarray}
%for the $S_{cc}^{++}(4100)$. 
\section{Conclusions}

We have considered the possible decay modes of the doubly charmed molecules of \cite{exotic}, $R_{cc}^+(3970)$, $S_{cc}^+(4100)$ and $S_{cc}^{++}(4100)$, and evaluated partial decay widths to $DD_{(s)}\pi$ and $DD_{(s)}\gamma$. We find that the main source of these decays come from the decay of a $D^*_{(s)}$ meson into $D_{(s)}\pi$ or $D_{(s)}\gamma$. The total widths of the doubly charmed molecules are $44\pm 12$ MeV for the $R_{cc}^+(3970)$ and $24\pm 8$ MeV for the $S_{cc}^{+}(4100)$ and $S_{cc}^{++}(4100)$. These decays are mediated by the exchange of one meson, vector or pseudoscalar, between the $D^*D^*_{(s)}$ pair of the molecule. The largest width comes from $\rho$, $\pi$ and $\omega$ exchange for the $R_{cc}(3970)$, and $K^*$, $K$, $J/\psi$ exchange for the  $S_{cc}(4100)$. %These mesons are a challenge for the experimentalist, since they are not $q\bar{q}$, have a pair of $cc$ and they are doubly charged. How to produce these mesons is a difficult question. Up to now, the only observed doubly charmed particle is the $\Xi_{cc}^+$, by its decays, $\lambda^+K^-\pi^+$ and $pD^+K^-$, however, the BABAR experiment didn\rq{}t find evidence for a $\Xi_{cc}^+$ in a search in $\Lambda_c^+ K^− \pi^+$ and $\Xi_c^0 \pi^+$ modes. The same for the BELLE experiment, without any evidence for a $\Xi_{cc}^+$ in the $\Lambda_c^+ K^- \pi^+$ mode. In the case of $e^+e^-$ collisions, Belle has produced doubly charm quarks in the final state, $J/\psi+c\bar{c}$, however the cross section is very small $\sigma(J/\psi+c\bar{c})=(0.87\pm 0.17)$ pb (see O. Seon), not being produced the $c\bar{c}$ pair without $J/\psi$ in the final state. More likely, they could be observed by the LHC in $pp$ collisions. 
%Maybe they can be produced together with doubly charmed baryons in a reaction like $\pi^- p\to \Xi^{+}_{cc}T^{-}$ or $K^- p\to \Xi^{++}_{cc}T^{--}_s$???.

These mesons are under challenge for experiments, since they are not $q\bar{q}$, having a pair of $cc$ and doubly charged. How to produce these mesons is a difficult question. Up to now, the only observed doubly charmed particle is the $\Xi_{cc}^+$, by its decays, $\Lambda^+K^-\pi^+$ and $pD^+K^-$ \cite{selex}, however, the BABAR experiment didn\rq{}t find evidence for a $\Xi_{cc}^+$ in a search in $\Lambda_c^+ K^− \pi^+$ and $\Xi_c^0 \pi^+$ modes \cite{babar}. The same for the BELLE experiment, without any evidence for a $\Xi_{cc}^+$ in the $\Lambda_c^+ K^- \pi^+$ mode \cite{belle}. In the case of $e^+e^-$ collisions, Belle has produced double charmed quarks in the final state, $J/\psi+c\bar{c}$, however the cross section is very small $\sigma(e^+e^-\to J/\psi+c\bar{c})=(0.74 \pm 0.08)$ pb and $J/\psi X_{\mathrm{non c\bar{c}}}$ cross section is $(0.43 \pm 0.09 \pm 0.09)$ pb \cite{ccjs}, not being able to produce the $c\bar{c}$ pair without $J/\psi$ in the final state. More likely, they could be observed by the LHC in $pp$ collisions or in JPARC, and being fortunated, the experimentalist could find both, the missed doubly charmed baryon and meson.

\begin{table}[H]
 \begin{center}
\begin{tabular}{lrrrrr}
\hline\hline
{\it State}& {\it Channel k}&$\Gamma^k$ [MeV] &{\it Channel j}&$\Gamma^k_j$ [MeV]&$\Gamma_{\mathrm{tot}}$ [MeV]\T\B\\\hline
$R^+_{cc}(3970)$&\multicolumn{5}{c}{{\it{Hadronic decays}}}\T\B\\
&$D^0D^{*+}$&$22\pm 6$&$D^0(D^{+}\pi^0)$&$7\pm 2$ & $44\pm 12$\T\B\\
&&&$D^0(D^{0}\pi^+)$& $15\pm 4$& \T\B\\
 &$D^+D^{*0}$&$22\pm 6$ &$D^+(D^0\pi^0)$&$14\pm 4$&\T\B\\
%&& &$D^+(D^+\pi^-)$&$-$\T\B\\
&\multicolumn{5}{c}{{\it{Radiative decays}}}\T\B\\
&$D^+D^{*0}$& &$D^+(D^0\gamma)$& $8\pm 2$&\T\B\\
&$D^0D^{*+}$&&$D^0(D^+\gamma)$&$0.4\pm 0.2$& \T\B\\
&& &$D^0D^+\gamma$&$(2\pm 1)\times10^{-3}$ &\T\B\\
 &&&$D^{*0}D^+\gamma$& $ (0.03\pm 0.01)\times10^{-3}$& \T\B\\
 % && &$(D^{+}\pi^-)D^+\gamma$ & \T\B\\
&& &$D^{*+}D^0\gamma$&$(0.5\pm 0.2)\times 10^{-3}$ &  \T\B\\
 \hline
 %\end{tabular}
 %\end{center}
% \caption{Coefficients in Eqs. ...}
% \label{tab:res2}
% \end{table}
% \begin{table}[htpb]
 %\begin{center}
%\begin{tabular}{lrrrr}
%\hline\hline
$S^+_{cc}(4100)$&\multicolumn{5}{c}{{\it{Hadronic decays}}}\T\B\\
&$D^+_sD^{*0}$&$12\pm 4$&$D^+_s(D^{0}\pi^0)$&$7\pm 2$&$24\pm 8$ \T\B\\
%&&&$D^+_s(D^{+}\pi^-)$&  \T\B\\
&$D^0D^{*+}_s$&$12\pm 4$&-&-&\T\B\\
&\multicolumn{5}{c}{{\it{Radiative decays}}}\T\B\\
&$D^0D^{*+}_s$&&$D^0(D^+_s\gamma)$&$11\pm 4$&\T\B\\
&$D^+_sD^{*0}$&&$D^+_s(D^0\gamma)$& $5\pm 2$&\T\B\\
&&&$D^0D^+_s\gamma$&$(2\pm 1)\times10^{-3}$ &  \T\B\\
&&&$D^{*0}D^+_s\gamma$&$(0.3\pm 0.1) \times 10^{-3}$ & \T\B\\
%& & & $(D^{+}\pi^-)D^+_s\gamma$& \T\B\\
%&& & $(D^{0}\gamma)D^+_s\gamma$&  \T\B\\
 &&&$D^{*+}_sD^0\gamma$&$(4\pm 1)\times 10^{-3}$ &\T\B\\
\hline
$S^{++}_{cc}(4100)$&\multicolumn{5}{c}{{\it{Hadronic decays}}}\T\B\\
&$D^+_sD^{*+}$&$12\pm 4$&$D^+_s(D^{+}\pi^0)$&$4\pm 1$& $24\pm 8$\T\B\\
&&&$D^+_s(D^{0}\pi^+)$&$8\pm 3$& \T\B\\
%&&&$D^+_s(D^{+}\pi^-)$&  \T\B\\
&$D^+D^{*+}_s$&$12\pm 4$&-&-&\\
&\multicolumn{5}{c}{{\it{Radiative decays}}}\T\B\\
&$D^+D^{*+}_s$&&$D^+(D^+_s\gamma)$&$11\pm 4$&\T\B\\
&$D^+_sD^{*+}$&&$D^+_s(D^+\gamma)$& $0.2\pm 0.1$&\T\B\\
&&&$D^+D^+_s\gamma$&$(1.3\pm 0.1)\times10^{-4}$ & \T\B\\
&&&$D^{*+}D^+_s\gamma$&$(0.3\pm 0.1)\times 10^{-3}$ & \T\B\\
%& & & $(D^{+}\pi^-)D^+_s\gamma$& \T\B\\
%&& & $(D^{0}\gamma)D^+_s\gamma$&  \T\B\\
 &&&$D^{*+}_sD^+\gamma$&$(0.3\pm 0.1)\times 10^{-3}$ &\T\B\\
 \hline\hline
\end{tabular}
\end{center}
\caption{Total and partial decay widths of the different decay modes of the doubly charmed states. }
\label{tab:res1}
\end{table} 

\begin{table}[H]
 \begin{center}
\begin{tabular}{lrr}
\hline\hline
{\it State}& {\it Intermediate meson}& $\Gamma_k$ [MeV]\T\B\\
\hline
$R_{cc}^+(3970)$&$\rho$ & $15.2$\T\B\\
&$\pi$ &$7.2$ \T\B\\
&$\omega$ & $1.7$\T\B\\
&$J/\psi$ & $0.6$\T\B\\
&$\eta$ &$0.14$ \T\B\\
&$\eta_c$ & $0.07$\T\B\\
&$\eta\rq{}$ & $0.018$\T\B\\
\hline
&$\rho+\pi$& $30.0$\T\B\\
&$\rho+\omega$&$7.0$\T\B\\
&$\pi+\omega$&$6.0$\T\B\\
\hline
$S^{+(+)}_{cc}(4100)$&$K^*$ &$ 15.0$\T\B\\
&$K$ & $4.3$\T\B\\
&$J/\psi$ & $1.7$\T\B\\
&$\eta$ &$0.4$ \T\B\\
&$\eta\rq{}$ &$0.2$ \T\B\\
&$\eta_c$ & $0.19$\T\B\\
\hline
&$K^*+K$&$24.9$\T\B\\
&$K^*+J/\psi$&$9.8$\T\B\\
&$J/\psi+K$&$4.2$\T\B\\
\hline
 \end{tabular}
\end{center}
\caption{Decay width obtained from the diagrams in Fig. \ref{fig:radocharm11} for one meson exchanged.}
\label{tab:par1}
\end{table}

\appendix
\section{Appendix}
\subsection{Pole positions of the dynamically generated XYZ using two different prescriptions for $g=m_V/(2f)$ in Eqs. (\ref{lcont}) and (\ref{l3V}).}

The pole positions and coupling constants in the hidden charm sector using $\alpha_h=-1.4$ and $g_l=m_\rho/(2 f_\pi)$ or $\alpha_h=-1.27$ and $g_h=m_{D^*}/(2f_D)$ in the $H-H$ channels. As one can observe, the use of a different strength $g_h$ for these channels is compensated by small changes in $\alpha_h$ and the strongest coupling to the main channel, $D^*\bar{D}^*$ or $D^*_s\bar{D}^*_s$ barely changes.

\begin{table}[H]
\begin{center}
\begin{tabular}{lrrrr}
\hline
&\multicolumn{3}{c}{$|g_R|$[MeV] }&($g_l$)\T\B\\
\hline
$\sqrt{s_0}$[MeV]&$3936 +i\,6 $&$3940+i\,0.0 $&$3921 +i\,30$&$4174 +i\,97$\T\B\\
$\alpha_h=-1.4$&$I=0;J=0$&$I=0;J=1$&$I=0;J=2$&$I=0;J=2$\T\B\\
\hline
$D^*\bar{D}^*$&$18700$&$18260$&$20600$&$2250$\T\B\\
$D_s^*\bar{D}^*_s$&$9900$&$9900$&$9050$&$20400$\T\B\\
$K^*\bar{K}^*$&$4$&$40$&$10$&$100$\T\B\\
$\rho\rho$&$50$&$0$&$100$&$90$\T\B\\
$\omega\omega$&$1400$&$0$&$2800$&$2970$\T\B\\
$\phi\phi$&$710$&$0$&$1600$&$3800$\T\B\\
$J/\psi J/\psi$&$390$&$0$&$2400$&$2400$\T\B\\
$\omega J/\psi$&$1600$&$0$&$4500$&$3400$\T\B\\
$\phi J/\psi$&$490$&$0$&$1900$&$6900$\T\B\\
$\omega \phi$&$60$&$0$&$190$&$2100$\T\B\\
\hline
&\multicolumn{3}{c}{$|g_R|$[MeV] }&($g_h$)\T\B\\\hline
$\sqrt{s_0}$[MeV]&$3950 +i\,12$&$3955 +i\,0.0$&$3922 +i\,24$&$4161 +i\,50$\T\B\\
$\alpha_h=-1.27$&$I=0;J=0$&$I=0;J=1$&$I=0;J=2$&$I=0;J=2$\T\B\\
\hline
$D^*\bar{D}^*$&$18000$&$17200$&$21000$&$900$\T\B\\
$D_s^*\bar{D}^*_s$&$10500$&$11000$&$5200$&$19500$\T\B\\
$K^*\bar{K}^*$&$30$&$80$&$50$&$90$\T\B\\
$\rho\rho$&$60$&$0$&$80$&$50$\T\B\\
$\omega\omega$&$1700$&$0$&$2300$&$1700$\T\B\\
$\phi\phi$&$1400$&$0$&$1800$&$2800$\T\B\\
$J/\psi J/\psi$&$390$&$0$&$3200$&$4450$\T\B\\
$\omega J/\psi$&$1700$&$0$&$3500$&$1800$\T\B\\
$\phi J/\psi$&$1400$&$0$&$2800$&$5500$\T\B\\
$\omega \phi$&$440$&$0$&$830$&$1900$\T\B\\
\hline
\hline
\end{tabular}
\end{center}
\caption{Pole positions and couplings $g_R$ to the different channels in two cases 1) using $g=m_\rho/(2f_\pi)$ in all channels, or 2) use of $g_h=m_{D^*}/(2f_D)$ in the channels where heavy mesons are involved, for Isospin$=0$.}
\label{xyza}
\end{table}

\begin{table}[H]
\begin{center}
\begin{tabular}{lrr}
\hline
&\multicolumn{2}{c}{$|g_R|$[MeV]}\T\B\\\hline
$\sqrt{s_0}$[MeV]&$3969 +i\,140 $&$3924 +i\,70$\T\B\\ 
$I=1;J=2$&\centering{( $g_l$)}&( $g_h$)\T\B\\\hline
$D^*\bar{D}^*$&$20500$&$20560$\T\B\\
$K^*\bar{K}^*$&$190$&$150$\T\B\\
$\rho\rho$&$0$&$0$\T\B\\
$\rho\omega$&$5000$&$3600$\T\B\\
$\rho J/\psi$&$8700$&$6200$\T\B\\
$\rho\phi$ &$3700$&$2600$\T\B\\\hline\hline
\end{tabular}
\end{center}
\caption{Pole positions and couplings $g_R$ to the different channels in two cases 1) using $g=m_\rho/(2f_\pi)$ in all channels, or 2) use of $g_h=m_{D^*}/(2f_D)$ in the channels where heavy mesons are involved, for Isospin$=1$.}
\label{xyzb}
\end{table}
\subsection{Decay of $R_c(S_c)$ to $DD^*$: Coefficients in Eq. (\ref{eq:int3}).}
The coefficients that survives in Eq. (\ref{eq:int3}),
\begin{eqnarray}
&&b=\frac{1}{16\pi^2}\int^1_0 dx\int^x_0 dy \frac{(2-y) x}{s(M_l)+i\eps}\nonumber\\
&&d=\frac{1}{16\pi^2}\int^1_0 dx\int^x_0 dy \frac{(y-2) y}{s(M_l)+i\eps}\nonumber\\
&&e= \frac{i}{(2 \pi)^3}f(P,k,M_l)+\frac{k^2_3}{16 \pi^2} \int^1_0 dx \int^x_0 dy \frac{y^2}{s(M_l)+i \eps}\nonumber\\
&&A=\frac{1}{P_0^2}\lbrace \frac{2i}{(2 \pi)^3} f_1(P,k,M_l)+\frac{1}{8\pi^2} \int_0^1dx\int_0^x dy\frac{k^2_3 y^2+k^2_0 y^2-2 P_0 k_0 x y+xP^2_0}{s(M_l)+i \eps}\rbrace\nonumber\\
%&&B=\frac{1}{16\pi^2}\int^1_0dx\int^x_0dy\frac{(2y-1)x}{s(M_l)+i\eps}\nonumber\\
&&C=\frac{1}{8\pi^2}\int^1_0dx\int^x_0 dy\frac{(x-1)y}{s(M_l)+i\eps}\nonumber\\
%&&D=\frac{1}{16\pi^2}\int^1_0dx\int^x_0dy\frac{y(1-2y)}{s(M_l)+i\eps}\nonumber\\
&&E=-2e(M_l)\ ,\label{eq:coefss}
\end{eqnarray}
with
\begin{eqnarray}
s(M_l)=(P^2-m^2_2+m^2_1)x+(-2Pk-M^2_l+m^2_2+k^2) y -(Px-yk)^2-m^2_1\ .\label{eq:fein31}
\end{eqnarray}

The integrals in Eqs. (\ref{eq:f}), (\ref{eq:f1}), and (\ref{eq:coefss}),
\begin{eqnarray}
&&f(P,k,M_l)=\int \frac{dq_0\,dq\,dx\,q^4x^2}{(q^2-M^2_1+i\eps)((P-q)^2-M^2_2+i\eps)((P-q-k)^2-M^2_l+i\eps)}\nonumber\\
&&=-i 2\pi\int_0^{q_{\mathrm{max}}}  dq \int^1_{-1} dx q^4\,x^2\times\nonumber\\&&
\times\left(\frac{1}{2\omega_1(P_0-\omega_1+\omega_2-i\eps)(P_0-\omega_1-\omega_2+i\eps)(k^0-P^0+\omega_1-\omega_l+i\eps)(k^0-P^0+\omega_1+\omega_l-i\eps)}\right .\nonumber\\&&\times
\frac{1}{2\omega_2(P_0-\omega_1+\omega_2+i\eps)(P_0+\omega_1+\omega_2-i\eps)(k^0+\omega_2-\omega_l+i\eps)(k^0+\omega_2+\omega_l-i\eps)}\nonumber\\&&\times\left .
\frac{1}{2\omega_l(P_0-k_0+\omega_1+\omega_l-i\eps)(P_0-k_0-\omega_1+\omega_l+i\eps)(k^0+\omega_2-\omega_l-i\eps)(k^0-\omega_2-\omega_l+i\eps)}\right)\nonumber\\\label{coff1}
\end{eqnarray}
\begin{eqnarray}
&&f_1(P,k,M_l)= \int \frac{ dq_0\,dq\,dx\,q^2(q^2_0+q^2 x^2)}{(q^2-m^2_1) ((P-q)^2-m^2_2) ((k+q-P)^2-M^2_l)}\nonumber\\&&=
-i\pi \int^1_{-1}dx\int^{q_{\mathrm{max}}}_0
   q^2dq\times\nonumber\\&&\times
      \left(\frac{q^2 x^2 + 
         \omega_1^2}{ \omega_1( i \eps + P_0 - \omega_1- \omega_2) (-i \eps + P_0 - 
           \omega_1 + \omega_2) (i \eps + P_0 - k_0 - 
           \omega_1- \omega_l) (-i \eps+ P_0 - 
           k_0 - \omega + 
           \omega_l}\right .\nonumber\\&&- \frac{-P_0^2 - q^2 x^2 - 
         2 P_0 \omega_2 - 
         \omega_2^2}{ (-i\eps - P_0 + \omega_1- \omega_2)\omega_2(P_0 + \omega_1 + \omega_2) (-i \eps - k_0-
            \omega_2 + \omega_l) (k_0 + 
           \omega_2+ \omega_l)} \nonumber\\&&\left .- \frac{-P_0^2 - q^2 x^2 -
          k_0^2 - 2 P_0 \omega_l - 
         \omega_l^2 + 
         2 k_0 (P_0 + \omega_l)}{(i \eps - P_0 + 
           k_0- \omega_1- 
           \omega_l) (-i \eps - P_0 + k_0 + 
           \omega_1 - \omega_l) (i \eps + 
           k_0- \omega_2 - 
           \omega_l) (-i \eps + k_0+ 
           \omega_2 - \omega_l) \omega_l}\right)\nonumber\\\label{coff2}
\end{eqnarray}
with $\omega_{1,2}=\sqrt{q^2+m^2_{1,2}}$, $\omega_l=\sqrt{q^2+|\vec{k}|\,^2+2q |\vec{k}|x+M^2_l}$ and $x=cos\theta$.

The coefficients in Eq. (\ref{eq:p}),
\begin{eqnarray}
&&\tilde{B} =\frac{1}{16\pi^2}\int^1_0dx\int^x_0 dy \frac{(x-1)(2-y)}{s(m_l)+i\eps}\nonumber\\
&&\tilde{E}=e(m_l)
\end{eqnarray}
Where $e(m_l)$ and $s(m_l)$ stands for $e$ and $s$ in the formulas of Eqs. (\ref{eq:coefss}) and (\ref{eq:fein31}), but the mass of the vector meson exchanged, $M_l$, replaced by the pseudoscalar one, $m_l$.
\subsection{Direct decays of $R_c(S_c)$ into $DD\gamma$: coefficients for the amplitudes from the Feynmann diagrams depicted in Fig. \ref{fig:radocharm11} 1), 2) and 3).}

The coefficients in Eq. (\ref{eq:ta}),
\begin{eqnarray}
&&a_1=-i\,6\int_0^1dx\int_0^xdy\int_0^ydz\frac{d^4 q\rq{}}{(2\pi)^4}\frac{a_1^b\vec{q}\,\rq{}^2+a_1^c}{(q\rq{}^2+s)^4}=\frac{1}{32 \pi^2}\int_0^1dx\int_0^xdy\int_0^ydz\left(\frac{-3a_1^b}{s}+\frac{2a_1^c}{s^2}\right)\ ,\nonumber\\
&&a_1^b=-2 - \frac{5 z}{3}\ ,\nonumber\\
&&a_1^c=-z (- 
   2 (-1 + x - y) (1 + z) \vec{k}\cdot\vec{ p}_1+ (x^2 - 2 x (1 + y) + y (2 + y)) 
\vec{p}_1\,^2)\ ,\nonumber\\\\
%ib_1=&&6\int_0^1dx\int_0^xdy\int_0^ydz\frac{d^4 q\rq{}}{(2\pi)^4}\frac{b_1^c}{(q\rq{}^2+s)^4}\nonumber\\
%b_1^c=&&2(-2+x+2xy-2y^2+y(-5+z)+z)\nonumber\\\\
&&c_1=-i\,6\int_0^1dx\int_0^xdy\int_0^ydz\frac{d^4 q\rq{}}{(2\pi)^4}\frac{c_1^c}{(q\rq{}^2+s)^4}=\frac{1}{16 \pi^2}\int_0^1dx\int_0^xdy\int_0^ydz\frac{c_1^c}{s^2}\ ,\nonumber\\
&&c_1^c=2(-2+x-y)(-2+2x-2y+z)\ ,\nonumber\\\\
&&d_1=-i\,6\int_0^1dx\int_0^xdy\int_0^ydz\frac{d^4 q\rq{}}{(2\pi)^4}\frac{d_1^b\vec{q}\,\rq{}^2+d_1^c}{(q\rq{}^2+s)^4}=\frac{1}{32 \pi^2}\int_0^1dx\int_0^xdy\int_0^ydz\left(\frac{-3d_1^b}{s}+\frac{2d_1^c}{s^2}\right)\ ,\nonumber\\
&&d_1^b=\frac{5}{3} (-2 + x - y)\ ,\nonumber\\
&&d_1^c=(-2 + x - y) ( - 
   2 (-1 + x - y) (1 + z) \vec{k}\cdot
    \vec{p}_1+ (x^2 - 2 x (1 + y) + y (2 + y)) 
\vec{p}_1\,^2)\ ,\nonumber\\
\end{eqnarray}
and
\begin{eqnarray}
&&s_1=(p^2_1+M^2_1-m^2_3) x+(P^2-M^2_2-p^2_1+m^2_3)y+(-2Pk-M^2_3+M^2_2) z\nonumber\\&&-M^2_1-(p_1 x+(P-p_1) y -kz)^2\ .
\end{eqnarray}
Where we have taken the trace in the terms proportional to $q\rq{}_l\,q\rq{}_m$. Also, we have used the formulas $\int\frac{d^4 q}{(q^2+s)^4}=\frac{i\pi^2}{6s^2}$, $\int\frac{d^4 q \,q^2}{(q^2+s)^4}=\frac{i\pi^2}{3s}$ and $\int\frac{d^4 q \,\vec{q}\,^2}{(q^2+s)^4}=-\frac{i\pi^2}{4s}$.

The coefficients in Eq. (\ref{eq:tb}),
\begin{eqnarray}
&&a_2=-i\,6\int_0^1dx\int_0^xdy\int_0^ydz\frac{d^4 q\rq{}}{(2\pi)^4}\frac{q\rq{}^2}{(q\rq{}^2+s_2)^4}=
\frac{1}{8 \pi^2}\int_0^1dx\int_0^xdy\int_0^ydz \frac{1}{s_2}\ ,\\
%ib_2=&&6\int_0^1dx\int_0^xdy\int_0^ydz\frac{d^4 q\rq{}}{(2\pi)^4}\frac{B^c_2}{(q\rq{}^2+s_2)^4}=\frac{i}{16 \pi^2}\int_0^1dx\int_0^xdy\int_0^ydz\frac{B^c}{s_2^2}\nonumber\\
%B^c_2\equiv &&B^c(x_i,k,p_1)=4 (y - z) (2 - x + y - z)\\
&&c=-i\,6\int_0^1dx\int_0^xdy\int_0^ydz\frac{d^4 q\rq{}}{(2\pi)^4}\frac{C^c}{(q\rq{}^2+s_2)^4}=\frac{1}{16 \pi^2}\int_0^1dx\int_0^xdy\int_0^ydz\frac{c^c}{s_2^2}\ ,\nonumber\\
&&c^c_2=-4 (2 + x^2 + y^2 + y (3 - 2 z) - 3 z + z^2 + x (-3 - 2 y + 2 z))\ ,\nonumber\\\\\nonumber
&&d_2=0\ ,
\end{eqnarray}
and 
\begin{eqnarray}
&&s_2= (-M_2^2 + m_3^2 + P^2 - 
    p_1^2) y + ( M_2^2 - m_4^2 - P^2 + 2 k p_1 + 
    p_1^2) z + (M_1^2 - m_3^2 + p_1^2) x\nonumber\\&&- (p_1 x + (P - p_1) y + (k - P + p_1) z)^2-M_1^2  \ .\nonumber\\\end{eqnarray}
   The coefficient $a_3$ in Eq. (\ref{eq:tc}),
    \begin{eqnarray}
&&a_3=-i\,2\int_0^1dx\int_0^xdy\frac{d^4 q\rq{}}{(2\pi)^4}\frac{(2-y)}{(q\rq{}^2+s_3)^3}=\frac{1}{16 \pi^2}\int_0^1dx\int_0^xdy\frac{(2-y)}{s}\\
&&s_3=-M_1^2 + (M_1^2 - M_2^2 + P^2) x + (k^2 + M_2^2 - m_3^2 - P^2 + 2 k p_1 + 
    p_1^2) y \nonumber\\&&- (P x + (k - P + p_1) y)^2\ ,\nonumber\\
\end{eqnarray}
\subsection{Decay of $R_c(S_c)$ to $DD^*_{(s)}$: Coefficients $C_V$ and $C_P$ in Eqs. (\ref{eq:squ}) and (\ref{eq:squ1}).}
Coefficients needed in Eqs. (\ref{eq:squ}) and (\ref{eq:squ1}), $C_V$ and $C_P$, for the evaluation of the decay $R_{cc}\to DD^*_{(s)}$. 
\begin{table}[H]
 \begin{center}
\begin{tabular}{lrrrr|rrrr}
\hline\hline
$V_1$&$V_2$&$m_l$&$m$&$V$&$I$&$A$&$P_V$&$C_P$\T\B\\
\hline
$D^{*+}$&$D^{*0}$&$\pi^0$&$D^+$&$D^{*0}$&$-\frac{1}{\sqrt{2}}$&$-\frac{1}{2}$&$\frac{1}{\sqrt{2}}$&$\frac{1}{4}$\T\B\\
$D^{*+}$&$D^{*0}$&$\eta$&$D^+$&$D^{*0}$&$-\frac{1}{\sqrt{2}}$&$-\frac{1}{\sqrt{6}}$&$-\frac{1}{\sqrt{3}}$&$-\frac{1}{6}$\T\B\\
$D^{*+}$&$D^{*0}$&$\eta\rq{}$&$D^+$&$D^{*0}$&$-\frac{1}{\sqrt{2}}$&$-\frac{1}{2\sqrt{3}}$&$-\frac{1}{\sqrt{6}}$&$-\frac{1}{12}$\T\B\\
$D^{*+}$&$D^{*0}$&$\eta_c$&$D^+$&$D^{*0}$&$-\frac{1}{\sqrt{2}}$&$-\frac{1}{\sqrt{2}}$&$1$&$\frac{1}{2}$\T\B\\
$D^{*0}$&$D^{*+}$&$\pi^-$&$D^+$&$D^{*0}$&$\frac{1}{\sqrt{2}}$&$-\frac{1}{\sqrt{2}}$&$-1$&$\frac{1}{2}$\T\B\\
$D^{*0}$&$D^{*+}$&$\pi^0$&$D^0$&$D^{*+}$&$\frac{1}{\sqrt{2}}$&$\frac{1}{2}$&$-\frac{1}{\sqrt{2}}$&$-\frac{1}{4}$\T\B\\
$D^{*0}$&$D^{*+}$&$\eta$&$D^0$&$D^{*+}$&$\frac{1}{\sqrt{2}}$&$-\frac{1}{\sqrt{6}}$&$-\frac{1}{\sqrt{3}}$&$\frac{1}{6}$\T\B\\
$D^{*0}$&$D^{*+}$&$\eta\rq{}$&$D^0$&$D^{*+}$&$\frac{1}{\sqrt{2}}$&$-\frac{1}{2\sqrt{3}}$&$-\frac{1}{\sqrt{6}}$&$\frac{1}{12}$\T\B\\
$D^{*0}$&$D^{*+}$&$\eta_c$&$D^0$&$D^{*+}$&$\frac{1}{\sqrt{2}}$&$-\frac{1}{\sqrt{2}}$&$1$&$-\frac{1}{2}$\T\B\\
$D^{*+}$&$D^{*0}$&$\pi^+$&$D^0$&$D^{*+}$&$-\frac{1}{\sqrt{2}}$&$-\frac{1}{\sqrt{2}}$&$-1$&$-\frac{1}{2}$\T\B\\
\hline
\end{tabular}
\end{center}
\caption{Coefficient $C_P$ in Eq. (\ref{eq:squ}) for $R_{cc}^+$.}
\label{tab:coef1}
\end{table}

\begin{table}[H]
 \begin{center}
\begin{tabular}{lrrrr|rrrr}
\hline\hline
$V_1$&$V_2$&$V_l$&$m$&$V$&$I$&$A$&$P_V$&$C_V$\T\B\\
\hline
$D^{*+}$&$D^{*0}$&$\rho^0$&$D^+$&$D^{*0}$&$-\frac{1}{\sqrt{2}}$&$-\frac{1}{2}$&$\frac{1}{\sqrt{2}}$&$\frac{1}{4}$\T\B\\
$D^{*+}$&$D^{*0}$&$\omega$&$D^+$&$D^{*0}$&$-\frac{1}{\sqrt{2}}$&$\frac{1}{2}$&$\frac{1}{\sqrt{2}}$&$-\frac{1}{4}$\T\B\\
$D^{*+}$&$D^{*0}$&$J/\psi$&$D^+$&$D^{*0}$&$-\frac{1}{\sqrt{2}}$&$\frac{1}{\sqrt{2}}$&$-1$&$\frac{1}{2}$\T\B\\
$D^{*0}$&$D^{*+}$&$\rho^+$&$D^+$&$D^{*0}$&$\frac{1}{\sqrt{2}}$&$\frac{1}{\sqrt{2}}$&$1$&$\frac{1}{2}$\T\B\\
$D^{*0}$&$D^{*+}$&$\rho^0$&$D^0$&$D^{*+}$&$\frac{1}{\sqrt{2}}$&$\frac{1}{2}$&$-\frac{1}{\sqrt{2}}$&$-\frac{1}{4}$\T\B\\
$D^{*0}$&$D^{*+}$&$\omega$&$D^0$&$D^{*+}$&$\frac{1}{\sqrt{2}}$&$\frac{1}{2}$&$\frac{1}{\sqrt{2}}$&$\frac{1}{4}$\T\B\\
$D^{*0}$&$D^{*+}$&$J/\psi$&$D^0$&$D^{*+}$&$\frac{1}{\sqrt{2}}$&$\frac{1}{\sqrt{2}}$&$-1$&$-\frac{1}{2}$\T\B\\
$D^{*+}$&$D^{*0}$&$\rho^-$&$D^0$&$D^{*+}$&$-\frac{1}{\sqrt{2}}$&$\frac{1}{\sqrt{2}}$&$1$&$-\frac{1}{2}$\T\B\\
\hline
\end{tabular}
\end{center}
\caption{Coefficient $C_V$ in Eq. (\ref{eq:squ}) for $R_{cc}^+$.}
\label{tab:coef2}
\end{table} 

\begin{table}[H]
 \begin{center}
\begin{tabular}{lrrrr|rrrr}
\hline\hline
$V_1$&$V_2$&$m_l$&$m$&$V$&$I$&$A$&$P_V$&$C_P$\T\B\\
\hline
$D^{*+}_s$&$D^{*0}$&$\eta$&$D^+_s$&$D^{*0}$&$-1$&$-\frac{1}{\sqrt{6}}$&$\frac{1}{\sqrt{3}}$&$\frac{1}{3\sqrt{2}}$\T\B\\
$D^{*+}_s$&$D^{*0}$&$\eta\rq{}$&$D^+_s$&$D^{*0}$&$-1$&$-\frac{1}{2\sqrt{3}}$&$-\sqrt{\frac{2}{3}}$&$-\frac{1}{3\sqrt{2}}$\T\B\\
$D^{*+}_s$&$D^{*0}$&$\eta_c$&$D^+_s$&$D^{*0}$&$-1$&$-\frac{1}{\sqrt{2}}$&$1$&$\frac{1}{\sqrt{2}}$\T\B\\
$D^{*0}$&$D^{*+}_s$&$K^-$&$D^+_s$&$D^{*0}$&$-1$&$-\frac{1}{\sqrt{2}}$&$-1$&$-\frac{1}{\sqrt{2}}$\T\B\\
$D^{*0}$&$D^{*+}_s$&$\eta$&$D^0$&$D^{*+}_s$&$-1$&$\frac{1}{\sqrt{6}}$&$-\frac{1}{\sqrt{3}}$&$\frac{1}{3\sqrt{2}}$\T\B\\
$D^{*0}$&$D^{*+}_s$&$\eta\rq{}$&$D^0$&$D^{*+}_s$&$-1$&$-\frac{1}{\sqrt{3}}$&$-\frac{1}{\sqrt{6}}$&$-\frac{1}{3\sqrt{2}}$\T\B\\
$D^{*0}$&$D^{*+}_s$&$\eta_c$&$D^0$&$D^{*+}_s$&$-1$&$-\frac{1}{\sqrt{2}}$&$1$&$\frac{1}{\sqrt{2}}$\T\B\\
$D^{*+}_s$&$D^{*0}$&$K^+$&$D^0$&$D^{*+}_s$&$-1$&$-\frac{1}{\sqrt{2}}$&$-1$&$-\frac{1}{\sqrt{2}}$\T\B\\
\hline
\end{tabular}
\end{center}
\caption{Coefficient $C_P$ in Eq. (\ref{eq:squ}) for $S_{cc}^+$.}
\label{tab:coef3}
\end{table}

\begin{table}[H]
 \begin{center}
\begin{tabular}{lrrrr|rrrr}
\hline\hline
$V_1$&$V_2$&$V_l$&$m$&$V$&$I$&$A$&$P_V$&$C_V$\T\B\\
\hline
$D^{*+}_s$&$D^{*0}$&$J/\psi$&$D^+_s$&$D^{*0}$&$-1$&$\frac{1}{\sqrt{2}}$&$-1$&$\frac{1}{\sqrt{2}}$\T\B\\
$D^{*0}$&$D^{*+}_s$&$K^{*+}$&$D^+_s$&$D^{*0}$&$-1$&$\frac{1}{\sqrt{2}}$&$1$&$-\frac{1}{\sqrt{2}}$\T\B\\
$D^{*0}$&$D^{*+}_s$&$J/\psi$&$D^0$&$D^{*+}_s$&$-1$&$\frac{1}{\sqrt{2}}$&$-1$&$\frac{1}{\sqrt{2}}$\T\B\\
$D^{*0}$&$D^{*+}_s$&$K^{*-}$&$D^0$&$D^{*+}_s$&$-1$&$\frac{1}{\sqrt{2}}$&$-1$&$\frac{1}{\sqrt{2}}$\T\B\\
\hline
\end{tabular}
\end{center}
\caption{Coefficients $C_V$ in Eq. (\ref{eq:squ}) for $S_{cc}^+$. }
\label{tab:coef4}
\end{table} 

\begin{table}[H]
 \begin{center}
\begin{tabular}{lrrrr|rrrr}
\hline\hline
$V_1$&$V_2$&$m_l$&$m$&$V$&$I$&$A$&$P_V$&$C_P$\T\B\\
\hline
$D^{*+}_s$&$D^{*+}$&$\eta$&$D^+_s$&$D^{*+}$&$1$&$-\frac{1}{\sqrt{6}}$&$\frac{1}{\sqrt{3}}$&$-\frac{1}{3\sqrt{2}}$\T\B\\
$D^{*+}_s$&$D^{*+}$&$\eta\rq{}$&$D^+_s$&$D^{*+}$&$1$&$-\frac{1}{2\sqrt{3}}$&$-\sqrt{\frac{2}{3}}$&$\frac{1}{3\sqrt{2}}$\T\B\\
$D^{*+}_s$&$D^{*+}$&$\eta_c$&$D^+_s$&$D^{*+}$&$1$&$-\frac{1}{\sqrt{2}}$&$1$&$-\frac{1}{\sqrt{2}}$\T\B\\
$D^{*+}$&$D^{*+}_s$&$\bar{K}^0$&$D^+_s$&$D^{*+}$&$1$&$-\frac{1}{\sqrt{2}}$&$-1$&$\frac{1}{\sqrt{2}}$\T\B\\
$D^{*+}$&$D^{*+}_s$&$\eta$&$D^+$&$D^{*+}_s$&$1$&$\frac{1}{\sqrt{6}}$&$-\frac{1}{\sqrt{3}}$&$-\frac{1}{3\sqrt{2}}$\T\B\\
$D^{*+}$&$D^{*+}_s$&$\eta\rq{}$&$D^+$&$D^{*+}_s$&$1$&$-\frac{1}{\sqrt{3}}$&$-\frac{1}{\sqrt{6}}$&$\frac{1}{3\sqrt{2}}$\T\B\\
$D^{*+}$&$D^{*+}_s$&$\eta_c$&$D^+$&$D^{*+}_s$&$1$&$-\frac{1}{\sqrt{2}}$&$1$&$-\frac{1}{\sqrt{2}}$\T\B\\
$D^{*+}_s$&$D^{*+}$&$K^0$&$D^+$&$D^{*+}_s$&$1$&$-\frac{1}{\sqrt{2}}$&$-1$&$\frac{1}{\sqrt{2}}$\T\B\\
\hline
\end{tabular}
\end{center}
\caption{Coefficient $C_P$ in Eq. (\ref{eq:squ}) for $S_{cc}^{++}$.}
\label{tab:coef5}
\end{table}

\begin{table}[H]
 \begin{center}
\begin{tabular}{lrrrr|rrrr}
\hline\hline
$V_1$&$V_2$&$V_l$&$m$&$V$&$I$&$A$&$P_V$&$C_V$\T\B\\
\hline
$D^{*+}_s$&$D^{*+}$&$J/\psi$&$D^+_s$&$D^{*+}$&$1$&$\frac{1}{\sqrt{2}}$&$-1$&$-\frac{1}{\sqrt{2}}$\T\B\\
$D^{*+}$&$D^{*+}_s$&$\bar{K}^{*0}$&$D^+_s$&$D^{*+}$&$1$&$\frac{1}{\sqrt{2}}$&$1$&$\frac{1}{\sqrt{2}}$\T\B\\
$D^{*+}$&$D^{*+}_s$&$J/\psi$&$D^+$&$D^{*+}_s$&$1$&$\frac{1}{\sqrt{2}}$&$-1$&$-\frac{1}{\sqrt{2}}$\T\B\\
$D^{*+}_s$&$D^{*+}$&$K^{*0}$&$D^+$&$D^{*+}_s$&$1$&$\frac{1}{\sqrt{2}}$&$1$&$\frac{1}{\sqrt{2}}$\T\B\\
\hline
\end{tabular}
\end{center}
\caption{Coefficient $C_V$ in Eq. (\ref{eq:squ}) for $S_{cc}^{++}$. }
\label{tab:coef6}
\end{table} 
\subsection{Direct decay of $R_{c}(S_{c})$ to $DD\gamma$: Coefficients $ {\cal A}$, ${\cal B}$ and ${\cal C}$ in Eqs. (\ref{eq:t2f}), (\ref{eq:coefs}) and (\ref{eq:coefs2}).}

Coefficients needed for the evaluation of the diagrams depicted in Fig. \ref{fig:radocharm11}) $1)-3)$, $ {\cal A}$, ${\cal B}$ and ${\cal C}$ in Eqs. (\ref{eq:t2f}), (\ref{eq:coefs}) and (\ref{eq:coefs2}).

\begin{table}[H]
 \begin{center}
\begin{tabular}{lrrrrrr|rrrrrr}
\hline\hline
$V_1$&$V_2$&$V_3$&$m_3$&$m_1$&$m_2$&$V_l$&${\cal P}$\T\B&$P_{V_1}$&$P_{V_2}$&$V_3$&$I$&${\cal A}$\\
\hline
$D^{*0}$&$D^{*+}$&$D^{*+}$&$\pi^0$&$D^0$&$D^+$&\T\B$\rho^0$&$\frac{1}{\sqrt{2}}$&$-\frac{1}{\sqrt{2}}$&$-\frac{1}{\sqrt{2}}$&$\frac{1}{\sqrt{2}}$&$\pm\frac{1}{\sqrt{2}}$&$\T\B\frac{1}{4\sqrt{2}}$\\
&&&$\eta$&&&&&$-\frac{1}{\sqrt{3}}$&$\frac{1}{\sqrt{3}}$&&&\T\B$-\frac{1}{6\sqrt{2}}$\\
&&&$\eta\rq{}$&&&&&$-\frac{1}{\sqrt{6}}$&$\frac{1}{\sqrt{6}}$\T\B&&&$-\frac{1}{12\sqrt{2}}$\\
&&&$\eta_c$&&&&&$1$&$-1$&&&$-\frac{1}{2\sqrt{2}}$\T\B\\
\hline
&&&$\pi^0$&&&$\omega$&$\frac{1}{3\sqrt{2}}$&$-\frac{1}{\sqrt{2}}$&$-\frac{1}{\sqrt{2}}$&\T\B$-\frac{1}{\sqrt{2}}$&$\pm\frac{1}{\sqrt{2}}$&$-\frac{1}{12\sqrt{2}}$\T\B\\
&&&$\eta$&&&&&$-\frac{1}{\sqrt{3}}$&$\frac{1}{\sqrt{3}}$&&&$\frac{1}{18\sqrt{2}}$\T\B\\
&&&$\eta\rq{}$&&&&&$-\frac{1}{\sqrt{6}}$&$\frac{1}{\sqrt{6}}$&&&$\frac{1}{36\sqrt{2}}$\T\B\\
&&&$\eta_c$&&&&&$1$&$-1$&&&$\frac{1}{6\sqrt{2}}$\T\B\\
\hline
&&&$\pi^0$&&&$J/\psi$&$\frac{2}{3}$&\T\B$-\frac{1}{\sqrt{2}}$&$-\frac{1}{\sqrt{2}}$&$1$&$\pm\frac{1}{\sqrt{2}}$&$\frac{1}{3\sqrt{2}}$\T\B\\
&&&$\eta$&&&&&$-\frac{1}{\sqrt{3}}$&$\frac{1}{\sqrt{3}}$&&&$-\frac{\sqrt{2}}{9}$\T\B\\
&&&$\eta\rq{}$&&&&&$-\frac{1}{\sqrt{6}}$&$\frac{1}{\sqrt{6}}$&&&$-\frac{1}{9\sqrt{2}}$\T\B\\
&&&$\eta_c$&&&&&$1$&$-1$&&&$-\frac{\sqrt{2}}{3}$\T\B\\
\hline
$D^{*0}$&$D^{*+}$&$D^{*+}$&$\pi^-$&$D^+$&$D^0$&$\rho^0$&$\frac{1}{\sqrt{2}}$&$-1$&$1$&$\frac{1}{\sqrt{2}}$&&$-\frac{1}{2\sqrt{2}}$\T\B\\
&&&&&&$\omega$&$\frac{1}{3\sqrt{2}}$&&&$-\frac{1}{\sqrt{2}}$&&$\frac{1}{6\sqrt{2}}$\T\B\\
&&&&&&$J/\psi$&$\frac{2}{3}$&&&$1$&&$-\frac{\sqrt{2}}{3}$\T\B\\\hline
%$D^{*0}$&$D^{*+}$&$D^{*+}$&$\pi^-$&$D^+$&$D^0$&$\rho^0$&$\frac{1}{\sqrt{2}}$&$-1$&$1$&$\frac{1}{\sqrt{2}}$&&$-\frac{1}{2\sqrt{2}}$\\
%&&&&&&$\omega$&$\frac{1}{3\sqrt{2}}$&&&$-\frac{1}{\sqrt{2}}$&&$\frac{1}{6\sqrt{2}}$\\
%&&&&&&$J/\psi$&$\frac{2}{3}$&&&$1$&&$-\frac{\sqrt{2}}{3}$\\\hline
\hline
\end{tabular}
\end{center}
\caption{Coefficient ${\cal A}$ in Eqs. (\ref{eq:t2f}), (\ref{eq:coefs}) and (\ref{eq:coefs2}) for $R_{cc}^{+}$.}
\label{tab:coef7}
\end{table} 
\begin{table}[H]
 \begin{center}
\begin{tabular}{lrrrrrr|rrrrrr}
\hline\hline
$V_1$&$V_2$&$V_3$&$m_3$&$m_1$&$m_2$&$V_l$&${\cal P}$\T\B&$P_{V_1}$&$P_{V_2}$&$V_3$&$I$&${\cal A}$\\
\hline
$D^{*0}$&$D^{*+}_s$&$D^{*+}_s$&$K^-$&$D^+_s$&$D^0$&$\phi$&$-\frac{1}{3}$&$-1$&$1$&$-1$&$-1$&\T\B$\frac{1}{3}$\\
&&&&&&$J/\psi$&$\frac{2}{3}$&&&$1$&&$\frac{2}{3}$\T\B\\\hline
$D^{*0}$&$D^{*+}_s$&$D^{*+}_s$&$\eta$&$D^0$&$D^+_s$&$\phi$&$-\frac{1}{3}$&$-\frac{1}{\sqrt{3}}$&$-\frac{1}{\sqrt{3}}$&$-1$&$-1$&$-\frac{1}{9}$\T\B\\
&&&$\eta\rq{}$&&&&&$-\frac{1}{\sqrt{6}}$&$\sqrt{\frac{2}{3}}$&&&$\frac{1}{9}$\T\B\\
&&&$\eta_c$&&&&&$1$&$-1$&&&$\frac{1}{3}$\T\B\\
\hline
&&&$\eta$&&&$J/\psi$&$\frac{2}{3}$&$-\frac{1}{\sqrt{3}}$&$-\frac{1}{\sqrt{3}}$&$1$&$-1$&$-\frac{2}{9}$\T\B\\
&&&$\eta\rq{}$&&&&&$-\frac{1}{\sqrt{6}}$&$\sqrt{\frac{2}{3}}$&&&$\frac{2}{9}$\T\B\\
&&&$\eta_c$&&&&&$1$&$-1$&&&$\frac{2}{3}$\T\B\\
\hline
\end{tabular}
\end{center}
\caption{Coefficient ${\cal A}$ in Eqs. (\ref{eq:t2f}), (\ref{eq:coefs}) and (\ref{eq:coefs2}) for $S_{cc}^{+}$.}
\label{tab:coef8}
\end{table} 
\begin{table}[H]
 \begin{center}
\begin{tabular}{lrrrrrr|rrrrrr}
\hline\hline
$V_1$&$V_2$&$V_3$&$m_3$&$m_1$&$m_2$&$V_l$&${\cal P}$\T\B&$P_{V_1}$&$P_{V_2}$&$V_3$&$I$&${\cal A}$\\
\hline
$D^{*+}$&$D^{*+}_s$&$D^{*+}_s$&$\bar{K}^0$&$D^+_s$&$D^+$&$\phi$&$-\frac{1}{3}$&$-1$&$1$&$-1$&$1$&$-\frac{1}{3}$\T\B\\
&&&&&&$J/\psi$&$\frac{2}{3}$&$-1$&$1$&$1$&$1$&$-\frac{2}{3}$\T\B\\
$D^{*+}$&$D^{*+}_s$&$D^{*+}_s$&$\eta$&$D^+$&$D^+_s$&$\phi$&$-\frac{1}{3}$&$-\frac{1}{\sqrt{3}}$&$-\frac{1}{\sqrt{3}}$&$-1$&$1$&$\frac{1}{9}$\T\B\\
&&&$\eta\rq{}$&&&&&$-\frac{1}{\sqrt{6}}$&$\sqrt{\frac{2}{3}}$&&&$-\frac{1}{9}$\T\B\\
&&&$\eta_c$&&&&&$1$&$-1$&&&$-\frac{1}{3}$\T\B\\
$D^{*+}$&$D^{*+}_s$&$D^{*+}_s$&$\eta$&$D^+$&$D^+_s$&\T\B$J/\psi$&$\frac{2}{3}$&$-\frac{1}{\sqrt{3}}$&$-\frac{1}{\sqrt{3}}$&$1$&$1$&$\frac{2}{9}$\T\B\\
&&&$\eta\rq{}$&&&&&$-\frac{1}{\sqrt{6}}$&$\sqrt{\frac{2}{3}}$\T\B&&&$-\frac{2}{9}$\\
&&&$\eta_c$&&&&&$1$&$-1$&&&$-\frac{2}{3}$\T\B\\
$D^{*+}_s$&$D^{*+}$&$D^{*+}$&$\bar{K}^0$&$D^+$&$D^+_s$&$\rho^0$&$\frac{1}{\sqrt{2}}$&$-1$&$1$&$\frac{1}{\sqrt{2}}$&$1$\T\B&$-\frac{1}{2}$\\
&&&&&&$\omega$&$\frac{1}{3\sqrt{2}}$&$-1$&$1$\T\B&$-\frac{1}{\sqrt{2}}$&$1$&$\frac{1}{6}$\\
&&&&&&$J/\psi$&$\frac{2}{3}$&$-1$&$1$&$1$&$1$&$-\frac{2}{3}$\T\B\\
%$D^{*+}_s$&$D^{*+}$&$D^{*+}$&$\bar{K}^0$&$D^+$&$D^+_s$&$J/\psi$&$\frac{1}{\sqrt{2}}$&$-1$&$1$&$\frac{1}{\sqrt{2}}$&$1$&\\
$D^{*+}_s$&$D^{*+}$&$D^{*+}$&$\eta$&$D^+_s$&$D^+$&$\rho$\T\B&$\frac{1}{\sqrt{2}}$&$\frac{1}{\sqrt{3}}$&$\frac{1}{\sqrt{3}}$&$\frac{1}{\sqrt{2}}$&$1$&$\frac{1}{6}$\T\B\\
&&&$\eta\rq{}$&&&&&$-\sqrt{\frac{2}{3}}$&$\frac{1}{\sqrt{6}}$&&&$-\frac{1}{6}$\T\B\\
&&&$\eta_c$&&&&&$1$&$-1$&&&$-\frac{1}{2}$\T\B\\
$D^{*+}_s$&$D^{*+}$&$D^{*+}$&$\eta$&$D^+_s$&$D^+$&\T\B$\omega$&$\frac{1}{3\sqrt{2}}$&$\frac{1}{\sqrt{3}}$&$\frac{1}{\sqrt{3}}$&$-\frac{1}{\sqrt{2}}$&$1$&$-\frac{1}{18}$\\
&&&$\eta\rq{}$&&&&&$-\sqrt{\frac{2}{3}}$&$\frac{1}{\sqrt{6}}$&&&$\frac{1}{18}$\T\B\\
&&&$\eta_c$&&&&&$1$&$-1$&&&$\frac{1}{6}$\T\B\\
$D^{*+}_s$&$D^{*+}$&$D^{*+}$&$\eta$&$D^+_s$&$D^+$&$J/\psi$&$\frac{2}{3}$&$\frac{1}{\sqrt{3}}$&$\frac{1}{\sqrt{3}}$&$1$&$1$&$\frac{2}{9}$\T\B\\
&&&$\eta\rq{}$&&&&&$-\sqrt{\frac{2}{3}}$&$\frac{1}{\sqrt{6}}$&&&$-\frac{2}{9}$\T\B\\
&&&$\eta_c$&&&&&$1$&$-1$&&&$-\frac{2}{3}$\T\B\\
\hline
\end{tabular}
\end{center}
\caption{Coefficient ${\cal A}$ in Eqs. (\ref{eq:t2f}), (\ref{eq:coefs}) and (\ref{eq:coefs2}) for $S_{cc}^{++}$.}
\label{tab:coef9}
\end{table}

\begin{table}[H]
 \begin{center}
\begin{tabular}{lrrrrrr|rrrrr}
\hline\hline
$V_1$&$V_2$&$m_3$&$m_4$&$m_1$&$m_2$&$V_l$&${\cal P}$&$P_{V_1}$&$P_{V_2}$&$P_{V_3}$&${\cal B}$\T\B\\\hline
$D^{*+}$&$D^{*0}$&$\pi^+$&$\pi^-$&$D^0$&$D^+$&$\rho^0$&$\frac{1}{\sqrt{2}}$&$-1$&$1$\T\B&$\sqrt{2}$&$\frac{1}{\sqrt{2}}$\\\hline
$D^{*+}_s$&$D^{*0}$&$K^+$&$K^-$&$D^0$&$D^+_s$&\T\B$\rho^0$&$\frac{1}{\sqrt{2}}$&$-1$&$1$&$\frac{1}{\sqrt{2}}$&$\frac{1}{2}$\\
&&&&&&$\omega$&$\frac{1}{3\sqrt{2}}$&&&$\frac{1}{\sqrt{2}}$\T\B&$\frac{1}{6}$\\
&&&&&&$\phi$&$-\frac{1}{3}$&&&$-1$&$\frac{1}{3}$\T\B\\\hline
\end{tabular}
\end{center}
\caption{Coefficient ${\cal B}$ in Eqs. (\ref{eq:t2f}), (\ref{eq:coefs}) and (\ref{eq:coefs2}).}
\label{tab:coef10}
\end{table} 

\begin{table}[H]
 \begin{center}
\begin{tabular}{lrrrr|rrrr}
\hline\hline
$V_1$&$V_2$&$m_3$&$m_1$&$m_2$&$V_{2P\gamma}$&$P_{V_1}$&$I$&${\cal C}$\T\B\\\hline
$D^{*0}$&$D^{*+}$&$\pi^+$&$D^+$&$D^0$&$-2$&$1$&$\pm \frac{1}{\sqrt{2}}$&$-\sqrt{2}$\T\B\\\hline
$D^{*+}$&$D^{*0}$&$\pi^-$&$D^0$&$D^+$&$1$&$1$&&$-\frac{1}{\sqrt{2}}$\T\B\\\hline
$D^{*+}$&$D^{*0}$&$\pi^0$&$D^+$&$D^0$&$\frac{1}{\sqrt{2}}$&$\frac{1}{\sqrt{2}}$&&\T\B$-\frac{1}{2\sqrt{2}}$\\
&&$\eta$&&&$-\frac{1}{\sqrt{3}}$&$\frac{1}{\sqrt{3}}$&&$\frac{1}{3\sqrt{2}}$\T\B\\
&&$\eta\rq{}$&&&$-\frac{1}{\sqrt{6}}$&$\frac{1}{\sqrt{6}}$&&$\frac{1}{6\sqrt{2}}$\T\B\\
&&$\eta_c$&&&$1$&$-1$&&$\frac{1}{\sqrt{2}}$\T\B\\\hline
$D^{*0}$&$D^{*+}_s$&$K^+$&$D^+_s$&$D^0$&$-2$&\T\B$1$&$-1$&$2$\\%\hline
%$D^{*0}$&$D^{*+}_s$&$\eta$&$D^0$&$D^+_s$&$0$&$-\frac{1}{\sqrt{3}}$&&$0$\\
%&&$\eta\rq{}$&&&&$\sqrt{\frac{2}{3}}$&&$0$\\
%&&$\eta_c$&&&&$-1$&&$0$\\
\hline
$D^{*+}_s$&$D^{*0}$&$K^-$&$D^0$&$D^+_s$&$1$&\T\B$1$&$-1$&$-1$\\\hline
%$D^{*+}_s$&$D^{*0}$&$\pi$&$D^+_s$&$D^0$&$0$&$\frac{1}{\sqrt{2}}$&$-1$&$0$\\
$D^{*+}_s$&$D^{*0}$&$\eta$&$D^+_s$&$D^0$&\T\B$\frac{1}{\sqrt{3}}$&$\frac{1}{\sqrt{3}}$&&$-\frac{1}{3}$\\
&&$\eta\rq{}$&&&$-\sqrt{\frac{2}{3}}$&\T\B$\frac{1}{\sqrt{6}}$&&$\frac{1}{3}$\\
&&$\eta_c$&&&$1$&\T\B$-1$&&$1$\\
\hline
%$D^{*0}$&$D^{*+}$&$\pi^0$&$D^0$&$D^+$&$0$&$-\frac{1}{\sqrt{2}}$&&$0$\\
%&&$\eta$&&&&$\frac{1}{\sqrt{3}}$&&$0$\\
%&&$\eta\rq{}$&&&&$\frac{1}{\sqrt{6}}$&&$0$\\
%&&$\eta_c$&&&&$-1$&&$0$\\
%\hline
\end{tabular}
\end{center}
\caption{Coefficient ${\cal C}$ in Eqs. (\ref{eq:t2f}), (\ref{eq:coefs}) and (\ref{eq:coefs2}). }
\label{tab:coef11}
\end{table}

\begin{table}[H]
 \begin{center}
\begin{tabular}{lrrrr|rrrr}
\hline\hline
$V_1$&$V_2$&$m_3$&$m_1$&$m_2$&$V_{2P\gamma}$&$P_{V_1}$&$I$&${\cal C}$\T\B\\\hline
$D^{*0}$&$D^{*+}$&$\pi^+$&$D^+$&$D^0$&$-2$&$1$&$\pm \frac{1}{\sqrt{2}}$&$-\sqrt{2}$\T\B\\\hline
$D^{*+}$&$D^{*0}$&$\pi^-$&$D^0$&$D^+$&$1$&$1$&&$-\frac{1}{\sqrt{2}}$\T\B\\\hline
$D^{*+}$&$D^{*0}$&$\pi^0$&$D^+$&$D^0$&$\frac{1}{\sqrt{2}}$&$\frac{1}{\sqrt{2}}$&&\T\B$-\frac{1}{2\sqrt{2}}$\\
&&$\eta$&&&$-\frac{1}{\sqrt{3}}$&$\frac{1}{\sqrt{3}}$&&$\frac{1}{3\sqrt{2}}$\T\B\\
&&$\eta\rq{}$&&&$-\frac{1}{\sqrt{6}}$&$\frac{1}{\sqrt{6}}$&&$\frac{1}{6\sqrt{2}}$\T\B\\
&&$\eta_c$&&&$1$&$-1$&&$\frac{1}{\sqrt{2}}$\T\B\\\hline
$D^{*0}$&$D^{*+}_s$&$K^+$&$D^+_s$&$D^0$&$-2$&\T\B$1$&$-1$&$2$\\%\hline
%$D^{*0}$&$D^{*+}_s$&$\eta$&$D^0$&$D^+_s$&$0$&$-\frac{1}{\sqrt{3}}$&&$0$\\
%&&$\eta\rq{}$&&&&$\sqrt{\frac{2}{3}}$&&$0$\\
%&&$\eta_c$&&&&$-1$&&$0$\\
\hline
$D^{*+}_s$&$D^{*0}$&$K^-$&$D^0$&$D^+_s$&$1$&\T\B$1$&$-1$&$-1$\\\hline
%$D^{*+}_s$&$D^{*0}$&$\pi$&$D^+_s$&$D^0$&$0$&$\frac{1}{\sqrt{2}}$&$-1$&$0$\\
$D^{*+}_s$&$D^{*0}$&$\eta$&$D^+_s$&$D^0$&\T\B$\frac{1}{\sqrt{3}}$&$\frac{1}{\sqrt{3}}$&$-1$&$-\frac{1}{3}$\\
&&$\eta\rq{}$&&&$-\sqrt{\frac{2}{3}}$&\T\B$\frac{1}{\sqrt{6}}$&&$\frac{1}{3}$\\
&&$\eta_c$&&&$1$&\T\B$-1$&&$1$\\
$D^{*+}_s$&$D^{*+}$&$\eta$&$D^+_s$&$D^+$&$\frac{1}{\sqrt{3}}$&$\frac{1}{\sqrt{3}}$&$1$&$\frac{1}{3}$\T\B\\
$D^{*+}_s$&$D^{*+}$&$\eta\rq{}$&$D^+_s$&$D^+$&$-\sqrt{\frac{2}{3}}$&$\frac{1}{\sqrt{6}}$&$1$&$-\frac{1}{3}$\T\B\\
$D^{*+}_s$&$D^{*+}$&$\eta_c$&$D^+_s$&$D^+$&$1$&$-1$&$1$&$-1$\T\B\\
$D^{*+}_s$&$D^{*+}$&$\bar{K}^0$&$D^+$&$D^+_s$&$-1$&$1$&$1$&$-1$\T\B\\
$D^{*+}$&$D^{*+}_s$&$\eta$&$D^+$&$D^+_s$&\T\B$-\frac{1}{\sqrt{3}}$&$-\frac{1}{\sqrt{3}}$&$1$&$\frac{1}{3}$\\
&&$\eta\rq{}$&&&$-\frac{1}{\sqrt{6}}$&\T\B$\sqrt{\frac{2}{3}}$&$1$&$-\frac{1}{3}$\\
&&$\eta_c$&&&$1$&\T\B$-1$&$1$&$-1$\\
$D^{*+}$&$D^{*+}_s$&$K^0$&$D^+_s$&$D^+$&$-1$&$1$&$1$&$-1$\T\B\\
\hline
%$D^{*0}$&$D^{*+}$&$\pi^0$&$D^0$&$D^+$&$0$&$-\frac{1}{\sqrt{2}}$&&$0$\\
%&&$\eta$&&&&$\frac{1}{\sqrt{3}}$&&$0$\\
%&&$\eta\rq{}$&&&&$\frac{1}{\sqrt{6}}$&&$0$\\
%&&$\eta_c$&&&&$-1$&&$0$\\
%\hline
\end{tabular}
\end{center}
\caption{Coefficient ${\cal C}$ in Eqs. (\ref{eq:t2f}), (\ref{eq:coefs}) and (\ref{eq:coefs2}). }
\label{tab:coef12}
\end{table}

\end{document}